# Anomalous bulk dealloying below the parting limit

Weiyue Zhou[1#,*], Hooman Gholamzadeh[2#], Lei Ding[3#], Kevin Daub[2], Mehdi Mosayebi[4], Travis Casagrande[4], Yingxin Zhu[3], Miaomiao Jin[5], Mark R. Daymond[2], Yang Yang[3,*], Michael P. Short[1,*], Suraj Y. Persaud[2,6,*]

1. *Department of Nuclear Science and Engineering, Massachusetts Institute of Technology, 77 Massachusetts Avenue, Cambridge, MA 02139, USA*
2. *Department of Mechanical and Materials Engineering, Smith Engineering, Queen's University, 60 Union St. W, Kingston, Ontario K7L 2N8, Canada*
3. *Department of Engineering Science and Mechanics, The Pennsylvania State University, University Park, PA 16802, USA*
4. *Canadian Centre for Electron Microscopy, McMaster University, 1280 Main St. W, Hamilton, Ontario L8S 4M1, Canada*
5. *Department of Nuclear Engineering, The Pennsylvania State University, University Park, PA 16802, USA*
6. *Department of Materials Science and Engineering, McMaster University, 1280 Main St. W, Hamilton, Ontario L8S 4M1, Canada*

* Email addresses of corresponding authors: wyzhou@mit.edu, yangyang@alum.mit.edu, hereiam@mit.edu, persas39@mcmaster.ca.
# These authors contributed equally.

**Abstract**

Dealloying has been extensively studied both as a corrosion degradation mechanism in structural materials, including those used in nuclear, aerospace, or marine environments, and as a versatile method to fabricate porous materials for catalysts and other functional applications. Classical dealloying theory in aqueous environments predicts a critical reactive-element concentration (parting limit) for continuous selective dissolution at temperatures where bulk diffusion does not dominate; this threshold is commonly reported around 50~60 at.%. Yet recent studies show that molten salt environments can generate extensive bulk dealloying below this threshold. Despite the importance of this anomalous dealloying behavior in many energy systems and electrochemical applications, its fundamental origin remains elusive. Here, we address this critical gap, revealing a grain boundary (GB)-assisted bulk dealloying mechanism. Using three-dimensional (3D) reconstruction of the dealloyed regions correlated with crystallographic and elemental analyses, we directly map the 3D GB–void architecture and reveal that diffusion-induced recrystallization (DIR) generates a high-density GB network, which then promotes molten-salt infiltration and can drive bulk dealloying far-below the conventionally reported parting limit, producing a distinctive morphology reminiscent of discontinuous precipitation (DP). Understanding this dynamic GB-void interplay is crucial for the prediction and control of dealloying in complex electrochemical environments.

## 1. Introduction

Alloying has been central to materials innovation since the Bronze Age, when small additions of alloying elements transformed soft, pure metals into stronger and more durable structural materials[1]. Modern alloy design extends this principle: by combining multiple principal elements and judicious trace additions, engineers can access an expansive compositional design space in which strength, ductility, corrosion resistance, and high-temperature stability can be tuned across wide ranges[2,3]. This compositional flexibility underpins many of the metallic materials that enable modern infrastructure and energy technologies, but it also creates a fundamental vulnerability in corrosive environments, where selective dissolution of the more reactive element in alloys can locally undermine the intended chemistry and microstructure. This process is known as dealloying, and has long been recognized as a corrosion mechanism that degrades structural alloys[4]. It has also recently become a powerful route to fabricate porous metals with large internal surface areas for catalysis, sensing, and electrochemical applications[5]. The ability to predict when selective dissolution remains confined to a surface region and when it propagates into the bulk is therefore central to both corrosion mitigation and the controlled synthesis of functional porosity[6].

Much of what is known about sustained dealloying derives from aqueous and liquid-metal environments. In aqueous systems, selective dissolution is often discussed in terms of a "parting limit". Below a critical concentration of the reactive element (commonly reported around 50-60 at.%), dissolution quickly self-arrests as the surface becomes enriched in the nobler element, which "passivates" the interface and blocks continued attack[7]. Related ideas also guide liquid-metal dealloying at elevated temperature, where dealloying can be harnessed to generate bicontinuous microstructures in alloy systems that are difficult to process in aqueous media[8]. Even across these diverse settings, a common expectation persists at intermediate homologous temperatures: when bulk diffusion does not dominate, the alloy must contain a sufficiently high concentration, and hence sufficient connectivity of the reactive element to sustain selective dissolution beyond the near-surface region[9].

This concentration constraint is not merely a theoretical detail. It limits the palette of alloys that can be transformed into porous materials without resorting to costly noble-rich compositions, and it bounds the conditions under which selective dissolution can be used for selective removal, recycling, or microstructure tailoring. At higher homologous temperatures, where bulk diffusion becomes dominant, the classical notion of a sharp parting limit becomes less useful because dealloying can proceed through bulk-diffusion-limited growth morphologies, such as "negative dendrite," that differ from bicontinuous percolation-based structures[10]. In practical applications, however, many molten-salt systems will be operated in an *intermediate regime*, just below temperatures where grain-boundary (GB)-dominated diffusion gives way to bulk-dominated mass transport. It is precisely in this regime that existing theory would predict self-limiting behavior for alloys below the critical reactive-element concentration.

Recent studies of molten-salt environments challenge the concept of a parting limit. Multiple reports show extensive void formation and salt-filled, interconnected porosity that penetrates deep into alloys[11–13] with reactive-element contents well below concentration thresholds inferred from aqueous (50-60 at.%[7]) and liquid-metal (possibly 40 at.%[8]) dealloying frameworks. These observations are difficult to reconcile with a purely bulk diffusion-controlled picture in which a stable solid–liquid interface recedes while isolated Kirkendall voids form internally[14]. Instead, the emerging evidence points to an evolving three-dimensional (3D) architecture in which void connectivity and salt infiltration become integral parts of the corrosion pathway[13]. Despite the practical importance of this "bulk dealloying" behavior for molten-salt-facing structural materials, and its tantalizing implications for using molten salts as a processing medium, the mechanistic origin of sustained selective dissolution below the expected critical concentration remains unresolved.

A key missing element is an explicit description of how microstructural evolution couples to transport and interface motion during molten-salt dealloying. In polycrystals, GBs can act as fast pathways for solute redistribution and as sites where the solid–liquid interface becomes locally unstable[13]. More importantly, under sustained chemical gradients, boundaries themselves can migrate and multiply through diffusion-driven processes, potentially creating new short-circuit pathways that did not exist in the initial microstructure[15,16]. If such boundary generation occurs concurrently with selective dissolution, the system could dynamically increase its internal pathway density, enabling salt penetration and reactive-element extraction even when the bulk composition falls below the percolation-based expectations that underpin the classical parting-limit concept[7].

Testing this idea requires direct, 3D mapping of the co-evolving GB network and void morphology, which is not accessible from conventional two-dimensional (2D) cross-sections alone. It is precisely this issue – typically static, 2D analysis of dynamically evolving 3D microstructures – which has collectively caused us to miss the mechanisms hidden in plain sight, in additional dimensions of space and time.

Here we reveal a GB-assisted bulk dealloying (GBABD) mechanism occurring in alloys immersed in a molten fluoride salt at intermediate homologous temperatures (0.58–0.65 of alloy melting points), in which diffusion-induced recrystallization (DIR) produces a high-density GB network that, in turn, promotes molten-salt infiltration and sustained selective dissolution in the bulk well below the expected parting limit. We study two model Ni–Cr alloys, Ni–22Cr (at.%) and Ni–16.6Cr (at.%), the latter selected specifically because its Cr content lies well below the face-centered-cubic (FCC) site-percolation limit that would be expected to suppress sustained dealloying in the intermediate-temperature regime. Using 3D focused ion beam-scanning electron microscopy (FIB-SEM) tomography correlated with crystallographic information from electron backscatter diffraction (EBSD) and chemical information from energy-dispersive X-ray spectroscopy (EDX), we directly reconstruct the 3D GB–void architecture that develops during exposure in FLiNaK + $EuF_3$ at intermediate homologous temperatures. The resulting morphologies display signatures reminiscent of diffusion-induced GB migration (DIGM) and DIR, alongside a distinctive tunnel-like porosity whose spatial organization evokes discontinuous precipitation (DP)–like colony growth.

By resolving how newly created and migrating GBs spatially correlate with interconnected, salt-filled voids, this work identifies a dynamic GB–void feedback loop as the enabling factor for bulk dealloying far below the predicted critical reactive-element concentration. Beyond explaining an outstanding inconsistency between molten-salt corrosion observations and established dealloying theory, the GBABD framework suggests a materials-design lever that is absent from classical parting-limit arguments: the ability of the microstructure to generate internal transport pathways during attack. Understanding and controlling this interplay is essential for predicting corrosion morphologies in molten salts, and for leveraging molten-salt dealloying as a scalable route to engineer porous metals and alloys.

## 2. Results

### 2.1 Identification of grain refinement and interconnected void network

**Fig. 1** and **Supplementary Fig. 1** present correlated 2D SEM and EBSD characterization of cross-sections of Ni-Cr alloys after 8 hours of exposure to molten FLiNaK + 5 wt.% $EuF_3$ at 700–800 °C. **Fig. 1a** illustrates the cross-sectioning workflow and highlights the region selected for analysis. Representative SEM and EBSD results from a Ni–22Cr alloy corroded at 700 °C are shown in **Fig. 1 b-c**, with the corresponding images for the other alloys and temperatures provided in **Supplementary Fig. 1**. In all cases, the corroded zone is characterized by a high density of voids, and spatially overlaps with a region of pronounced grain refinement, consistent with recrystallization. The co-localization of void formation and grain refinement indicates that void evolution and GB microstructure reorganization occur concurrently during molten-salt exposure.

Quantitative grain-size statistics (**Fig. 1d**) confirm this trend across all three conditions: the dealloyed regions exhibit a substantially smaller grain size than the underlying, unaffected material. The inverse pole figure maps along the sample normal (IPF-Z) in **Fig. 1c** and **Supplementary Fig. 1d–f** further show that the refined-grain zones coincide with the corroded regions and possess crystallographic orientations distinct from the parent grains beneath the dealloyed layer, supporting a recrystallization-based origin rather than simple subgrain subdivision.

Boundary-character analyses provide additional insight into the architecture of the refined region. High-angle GB maps (15°–55°) in **Supplementary Fig. 1g–i**, together with Σ3 maps in **Supplementary Fig. 1j–l**, show that the newly formed grains organize into domains separated by high-angle boundaries, within which a high density of Σ3 boundaries are present. This behavior is consistent across the three samples, as reflected by the quantitative twin-boundary density analysis (**Fig. 1e**). Finally, comparison of the surface microstructure before and after exposure (**Supplementary Fig. 2**) indicates that these recrystallized regions, observed in cross-section in **Fig. 1** and **Supplementary Fig. 1**, are typically associated with the original GB network, suggesting that pre-existing GBs act as preferred initiation sites for the coupled recrystallization and dealloying process.

The 2D cross-sectional SEM images in **Fig. 1b** resolve the distribution of voids and crevices within the recrystallized zone. SEM–EDX (**Supplementary Fig. 3**) detects salt species inside these voids, indicating that they are not isolated cavities, but rather part of an interconnected pore space that permits liquid-salt infiltration. This finding motivated a 3D reconstruction of both the void network and the GB network to determine how their evolution is coupled.

To capture a sufficiently large field of view while retaining the spatial resolution needed to resolve narrow void channels and boundary connectivity (two requirements that typically trade off), we employ FIB–SEM tomography. We further integrate EBSD on the cross-sections exposed during serial sectioning, allowing us to use the FIB–SEM stack to reconstruct the 3D void morphology while using the EBSD maps to identify grains and delineate GBs within the same coordinate system. Achieving robust correlation between these two datasets is non-trivial because it requires reliable image drift correction, feature extraction, correlation and classification, EBSD indexing over hundreds of slices, and accurate registration between morphological and crystallographic volumes. We therefore developed a dedicated workflow (Methods; Supplementary Information) that enables co-registered 3D reconstructions of void space and GB topology.

**Fig. 2** presents the resulting 3D reconstructions of the void spaces (**Fig. 2a, b**), voids + GBs (**Fig. 2c, d**), and GBs (**Fig. 2e, f**), respectively, confirming that the majority of voids form a connected network and directly resolving their spatial relationship with the evolving GB architecture. The corresponding **Supplementary Videos 1-6** provide an intuitive view of the 3D morphology and connectivity of the salt-accessible void network.

These observations are inconsistent with the conventional attribution of similar cross-sectional features to Kirkendall voiding[14]. Specifically, our 3D reconstructions show that the porosity is predominantly interconnected and salt-accessible, rather than a population of isolated vacancies condensing into closed voids as expected for a purely bulk interdiffusion-driven process.

### 2.2 Grain boundary evolution led by synergistic DIR and DIGM

The dealloyed regions shown in **Fig. 2** exhibit coupled, concurrent evolution via three mechanisms: selective dissolution creates and enlarges an infiltrating pore network, salt transport proceeds through this connected porosity, and concurrent GB restructuring (including the formation of new boundaries) continuously reshapes the pathways that sustain attack.

Before discussing the role of porosity, we first consider the origin of the pronounced grain refinement in the dealloyed region. Grain refinement can arise from several mechanisms, including conventional static recrystallization driven by deformation-induced stored strain energy[17], dynamic recrystallization during high-temperature deformation[18], continuous recrystallization through dislocation absorption[19], particle-stimulated nucleation[20], and shear-band- or twin-assisted recrystallization[21,22]. These mechanisms are unlikely to dominate here because the grain refinement occurs during corrosion exposure without externally imposed plastic deformation, and because the newly formed boundaries are spatially correlated with GBs, surface/GB junctions, and strong chemical redistribution.

Instead, the observations are more consistent with diffusion-induced recrystallization, or DIR[16]. DIR is a chemically driven recrystallization process observed in diffusion couples, where rapid compositional changes near interfaces or GBs trigger the nucleation of new grains, followed/coupled by growth through diffusion-induced GB migration, DIGM[23]. Several features of the present system support this interpretation: the process initiates near original GBs and GB-surface junctions, involves strong interdiffusion and selective dissolution, and produces a high density of high-angle GBs within the affected region. These features are consistent with prior descriptions of DIR/DIGM in chemically evolving diffusion systems[23,24].

Although DIR has been discussed mostly in alloying systems[25], the same physical framework can apply to dealloying. In both cases, a moving GB separates regions with different local compositions. The associated lattice misfit stresses ahead of the moving GB can generate coherency strain[26,27], which provides a driving force for GB migration[28] or, when the strain cannot be sufficiently relaxed by migration, for the nucleation of new grains[29]. In the present molten-salt dealloying system, preferential Cr dissolution and resulting Ni enrichment create strong local composition gradients near GBs and GB-surface junctions. These gradients can therefore promote DIGM and DIR even though the net chemical process is dealloying rather than alloying.

Once new grains form, they introduce additional high-angle GBs and increase the density of short-circuit diffusion pathways. Chemical redistribution can then continue along these newly generated boundaries, promoting further GB migration and additional recrystallization. This creates a positive feedback loop: selective dissolution drives chemical gradients, chemical gradients promote DIR/DIGM, and DIR/DIGM generates new GB pathways that accelerate Cr transport toward the molten salt. Thus, the dealloyed region dynamically creates new diffusion highways rather than relying only on the original GB network.

This mechanism also distinguishes the present process from conventional Kirkendall voiding[30]. Although the system contains unequal diffusion fluxes and a net loss of Cr, the resulting free volume does not appear primarily as isolated Kirkendall voids. Instead, it forms connected tunnels infiltrated by liquid salt. These tunnels bring the solid/liquid interface closer to Cr-rich regions, while DIR increases the GB area available for short-circuit diffusion. Together, these two processes enable rapid and spatially extensive dealloying beyond what would be expected from bulk diffusion, original GB diffusion, or isolated Kirkendall void formation alone.

### 2.3 GB-guided formation of void tunnels

We define a *dealloying colony* as the 3D volume enclosed by the advancing corrosion front (that is, the dealloyed domain bounded by the migrating reaction interface). At the colony scale, the dealloyed regions typically adopt a half-elliptical outline in cross-section (**Supplementary Fig. 4**), and a representative 3D example is shown in **Fig. 2a** (white dashed outline). EBSD in both 2D (**Fig. 1**) and 3D (**Fig. 2**) shows that the leading edge of each colony coincides with GBs, establishing that the corrosion front is GB-guided. Within this GB-guided architecture, the void tunnels in the DIR region are not incidental: their orientations, bends, and branching patterns are highly structured within each colony and therefore provide morphological "fingerprints" that constrain how the front advances and how salt-accessible porosity is generated.

A particularly diagnostic feature is the alignment of void tunnels relative to the local dealloying front. Near the front, tunnels commonly run either parallel or perpendicular to the migrating GB segment (black arrows in **Fig. 2a**), with occasional branching. The preferred orientation correlates with the local boundary geometry relative to the sample surface: when the GB segment is close to being perpendicular to the sample surface, tunnels more often remain parallel to the GB; when the GB segment is close to being parallel to the sample surface, tunnels more often develop approximately perpendicular to the GB. These non-random alignments, together with the occurrence of sharp bends and branching, indicate that tunnel growth is coupled to GB migration and the dealloying-front geometry, rather than arising from spatially random void nucleation.

The morphology also suggests a natural sequence. A recurring "early" configuration is a salt-filled tunnel that lies on a GB plane and runs parallel to it, consistent with surface-initiated, GB-confined channel growth. This configuration matches the initiation picture proposed in our previous work as "1D wormhole corrosion," in which salt-filled channels propagate along GBs and form rope-like, percolating structures that lie on GB planes[13]. In the present system, these GB-parallel wormholes provide a direct mechanism to create salt-accessible internal interfaces and to establish an interconnected network at the beginning of colony development. The central mechanistic question then becomes how these initially GB-parallel tunnels evolve into the front-perpendicular tunnel arrays that dominate in more developed regions.

The 3D reconstructions identify two recurring pathways for reorientation, both controlled by GB migration. The first pathway is illustrated schematically in **Fig. 3a** and supported by 2D and 3D evidence in **Fig. 3b, c** (see **Supplementary Video 7** for the 3D visualization). Here, a tunnel that originally lay along a GB remains geometrically parallel to the local boundary trace, but becomes embedded within a grain because the GB migrates away from it. In other words, the moving GB "shakes off" a pre-existing wormhole, leaving behind a tunnel that records a former GB position. A key feature of this scenario is that the detached tunnel can remain connected to the active GB at its tip, so that the tunnel no longer resides on the boundary plane but still participates in transport and reaction at the front.

The second pathway explains how a tunnel initially parallel to a GB can become perpendicular to it. This is captured schematically in **Fig. 3d**, with 2D and 3D evidence in **Fig. 3e, f (**see **Supplementary Video 8** for the 3D visualization**)**. This pathway arises when adjacent segments of a tortuous GB migrate at different rates. As the faster segment advances, the tunnel continues to grow along the moving boundary segment, while remaining connected to a neighboring, slower migrating segment. This differential motion bends the tunnel and produces a sharp, nearly 90° turn, yielding the characteristic "L"-shaped morphology. When the slower GB segment later advances and detaches from the tunnel portion to which it had remained parallel, the tunnel tip becomes effectively perpendicular to the GB. The abundance of "L"-shaped and more complex ("F"-like) tunnels within colonies (**Fig. 2**) is consistent with repeated differential-migration and progressive detachment events, rather than a single, abrupt reorientation step.

Once a tunnel tip becomes perpendicular to the migrating GB at the dealloying front, its subsequent evolution is consistent with a front-coupled growth mode. **Fig. 4a** schematically illustrates the geometric consequence: as the GB advances, it drags the tunnel tip forward, elongating the salt-filled tunnel in the wake of migration. This "dragging/elongation" mechanism is supported by the 2D cross-sectional SEM image in **Fig. 4b** and the corresponding 3D reconstruction in **Fig. 4c (**see **Supplementary Video 9** for the 3D visualization**)**, which show elongated, front-perpendicular tunnels that track the migration direction. In this stage, branching becomes a frequent additional signature. **Fig. 4d** schematically depicts a branching event that occurs during front-coupled elongation; the corresponding 2D SEM evidence and 3D reconstruction are shown in **Fig. 4e, f (**see **Supplementary Video 10** for the 3D visualization**).** The combination of quasi-parallel, front-perpendicular tunnel arrays and branching morphologies is strongly reminiscent of discontinuous precipitation (DP), in which a migrating boundary mediates solute redistribution and produces an array of needle- or plate-like product features that are roughly parallel to one another and oriented perpendicular to the migration direction[31–33]. Here, the resemblance is morphological and kinetic: the evolving front organizes an aligned product array, and branching is a common outcome of growth[34].

At the same time, the initiation and "product identity" differ from classical DP in a way that is essential for molten-salt dealloying. In DP, the secondary phase nucleates along GBs within the bulk, often driven by supersaturation following a temperature shift, and nucleation does not require any external surface contribution. In contrast, the salt-filled tunnels in the present system must ultimately be surface-connected to become salt-accessible, consistent with the GB-parallel wormhole initiation described above. Moreover, although the final observed structure is a void space filled with molten salt, the DP-like "second phase" is more naturally interpreted as a conceptual counterpart to the secondary phase in classical DP. Whereas solute influx in classical DP produces a retained solute-rich phase, Cr influx in the present system is driven by continuous Cr removal into the oxidizing molten salt at the migrating front. This salt-mediated removal both creates the chemical-potential gradient for Cr transport and leaves behind a salt-filled space that conceptually replaces the solute-rich product phase in classical DP. Thus, the observed tunnel serves as forensic evidence for a DP-like redistribution process in which the would-be Cr-rich phase is continuously dissolved as it is supplied.

**Supplementary Fig. 5** maps the specific subvolumes extracted from the full 3D reconstruction (**Fig. 2**) that are shown in **Figs. 3-4**, and provides rotated views of the same local structures to clarify their 3D geometry and orientation.

### 2.4 GB assisted bulk dealloying (GBABD)

The results thus far establish a consistent mechanism for how the coupled void–GB architecture in **Fig. 2** emerges during molten-salt dealloying, enabling bulk dealloying at compositions below the classical parting-limit expectation. This mechanism, which we name "GB assisted bulk dealloying (GBABD)," is summarized schematically in **Fig. 5** using a cross-sectional view.

From a coupled alloy–salt perspective, the Ni–Cr alloy is thermodynamically driven toward Cr removal under sufficiently oxidizing salt conditions: given effective pathways, Cr can dissolve into the molten salt as a soluble specie. In the early stage, pre-existing GBs act as short-circuit diffusion pathways, so Cr depletion and leaching proceed preferentially along GBs compared with the grain interior (**Fig. 5a**). The resulting compositional gradients promote GB migration and boundary generation through DIGM and DIR (**Fig. 5b**). This step is essential because it transforms the microstructure from one governed by a sparse, inherited GB network into one containing a dense and continuously evolving boundary network that can repeatedly renew transport pathways.

Surface-connected salt infiltration then converts these fast pathways into internal reaction interfaces. Wormholes initiate at GB–surface junctions and propagate along GB planes into the bulk, forming salt-filled channels that immediately create internal solid–liquid interface area (**Fig. 5c**). As GBs continue to migrate, wormholes can detach from the active GB plane through a "shake-off" process (**Fig. 5d**), leaving behind salt-filled tunnels that record former GB positions while remaining part of the connected pore network. The newly positioned migrating GB can then nucleate additional wormholes, followed by further shake-off events (**Fig. 5e**), progressively building a colony-scale architecture in which salt-accessible tunnels and newly generated GBs co-exist and co-evolve.

Once tunnels become oriented approximately perpendicular to the migrating GB segment at the dealloying front, continued front motion can elongate these tunnels in a DP-like manner (**Fig. 5f**). In this regime, the advancing GB effectively drags the tunnel tip forward, extending the salt-filled channel while maintaining access to the molten salt. Local variations in boundary geometry and migration rate, together with the presence of pre-existing tunnel segments, can generate branching during elongation (**Fig. 5g**), which increases connectivity and further amplifies internal solid–liquid interfacial area. Repetition of these coupled steps (**Fig. 5**) produces the complex, interconnected void–GB network observed by 3D imaging (**Fig. 5h**; compare with **Fig. 2**).

This framework also clarifies why bulk dealloying persists even when the bulk composition is below the predicted parting limit. Classical parting-limit arguments assume that sustained dissolution requires sufficient reactive-element connectivity beyond the dissolution interface[7]; otherwise, the surface becomes "passivated" by the noble metal atoms facilitated by surface diffusion and arrests dealloying. In contrast, GBABD continuously creates internal, salt-accessible solid–liquid interfaces (tunnels) and continuously renews short-circuit pathways (i.e., GBs) through DIR/DIGM; thus, a pre-existing percolating pathway of the reactive-element atoms is not a necessary condition for dealloying in this regime. Cr can be transported efficiently to tunnel surfaces via connected GBs and dissolve into the salt, while the remaining Ni-rich matrix persists transiently because bulk diffusion is comparatively slow at these intermediate homologous temperatures to allow long-range bulk transport. After a boundary segment passes, any residual Cr becomes kinetically trapped unless additional DIR generates new GB pathways that reopen short-circuit transport and re-enable salt access. In this sense, sustained bulk dealloying arises from a dynamic feedback loop between GB evolution and salt infiltration, rather than from the static percolation of the reactive element in the original alloy microstructure.

## 3. Discussions

In this work, we proposed a GBABD mechanism to explain the coupled tunnel–GB architecture and the sustained bulk dealloying observed below the classical parting limit during molten-salt corrosion. Through the creation and migration of these boundaries, the bulk material can undergo dealloying at temperatures where direct bulk dealloying via negative dendritic dealloying[10] is unfavorable, and percolation dealloying is inhibited due to low concentrations of more reactive elements (e.g. Cr of Ni-Cr facing molten fluoride salt)[7]. At the low concentrations demonstrated in the current study (down to 16.6 at. %, which is significantly below the FCC single-atom 3D site percolation threshold), the dealloying process generates an interconnected network of tunnels rather than forming Kirkendall voids.

A practical question is the temperature window in which GBABD operates or dominates. We use "intermediate homologous temperature" in the phenomenological sense associated with the Type B regime in Harrison's classification of GB diffusion[35,36], in which GBs provide fast short-circuit diffusion paths while lattice diffusion remains sufficient to couple the boundary to adjacent grain interiors. Sharp numerical boundaries are therefore difficult to generalize, because these transitions depend on alloy composition, salt redox potential, grain size, and competing kinetics. Operationally, GBABD requires that DIGM is active but remains below conditions where negative dendritic dealloying becomes dominant[10]. If DIGM is inactive, dealloying may be largely confined to GBs without extensive bulk penetration. Although DIGM is most often reported over $T/T_{MELT}$ = 0.25~0.65, this range should not be treated as a strict limit[25]. DIR likely imposes more stringent requirements than DIGM; therefore, GBABD variants that involve widespread new-grain formation are expected to occur at higher temperatures (or stronger chemical driving forces) than cases in which DIR is absent, all else being equal.

This framework suggests several testable predictions. Single crystals of alloys should still be capable of dealloying via GBABD if DIR nucleates new boundaries; however, without pre-existing

GBs, recrystallization will likely require stronger driving forces (for example, higher reactive-element content) to accumulate sufficient coherency strain for nucleation. In polycrystals, the original GB network can play a similar enabling role, reducing the extent of chemical modification required to initiate boundary-guided penetration. Conversely, at higher reactive-element concentrations, DIR-assisted boundary multiplication should become more prevalent, whereas at lower concentrations DIGM-dominated evolution may proceed with less boundary generation, assuming comparable redox conditions. When interfacial reactions are slow, coherency-strain accumulation may be hindered by relaxation associated with flux imbalance and Kirkendall-type effects; in such cases, isolated voiding becomes more likely unless negative dendritic dealloying takes over.

Several open questions remain. It is unclear whether GBABD has a composition threshold below which sustained bulk penetration cannot occur even with GB assistance. If such a threshold exists, it would not be governed by the surface-diffusion and geometric constraints that govern classical percolation dealloying. Instead, it would depend on the propensity for DIGM and DIR and on the ability to initiate and sustain surface-connected wormholes. Microstructural constraints could therefore impose an effective "GBABD parting limit:" stable pinning precipitates or the presence of a third major element that reduces coherency strain may suppress GB mobility and attenuate bulk penetration. If a coherency-strain-related threshold exists, it would most likely correspond to DIGM, because the corresponding threshold for DIR is expected to be higher. Given the importance of DIR and DIGM for facilitating selective dissolution, cold work and plastic deformation may also influence the existence or value of such a parting limit.

Mechanistically, while coherency strain is a natural driver for DIR in chemically modified regions, the role of dislocations and vacancy creation/annihilation under unequal atomic fluxes during bulk Kirkendall diffusion[37] remains an important open topic. DIR via dislocation polygonization, analogous to recrystallization in deformed solids, has also been proposed[23]. For DIGM, proposed drivers include coherency strain[28], GB Kirkendall effects coupled with GB dislocation climb[38], and gradient-energy corrections to chemical potentials[39], among others. Flux-imbalance–driven defect processes may contribute to both DIGM and DIR. Resolving their relative importance will require early-stage, defect-sensitive characterization within the DIR/DIGM zone[13] and quantitative modeling that links chemical gradients, defect kinetics, and boundary mobility.

Looking forward, GBABD reframes molten-salt dealloying as a dynamic feedback process: boundary evolution creates transport pathways and internal solid–liquid interfaces that, in turn, sustain selective dissolution in the bulk, even below the classical parting limit. More broadly, this mechanism offers a new interpretation of dealloying in an intermediate temperature regime, bridging percolation-controlled dealloying at lower temperatures[6] and lattice-diffusion-controlled dealloying at higher temperatures[10]. This perspective suggests practical levers for mitigation (controlling GB mobility through alloying/precipitate engineering and managing salt redox potential) and also points to an opportunity: using molten salts to engineer porous architectures in alloys that would be "non-dealloyable" under conventional parting-limit expectations.

## 4. Materials and Methods

### Materials

0.5 mm thick Ni-15Cr and Ni-20Cr (wt. %) model alloy sheets were manufactured by Sophisticated Alloys Inc. with a certified purity level of 99.95 wt. %. The corresponding alloy compositions in atomic percent are Ni-16.6Cr and Ni-22Cr, respectively, which will be the notation used to describe the alloys throughout this paper. Electron backscatter diffraction (EBSD) results show that all samples used in this study are single-phase, face-centered cubic (FCC) structures. Samples were cut into rectangular shapes with approximate dimensions of 12 × 10 mm. Prior to corrosion testing, all surfaces - including the two major and four minor faces - were mechanically polished using SiC paper of sequentially finer grit, followed by polishing with diamond suspension down to 1 μm finish, and finalized with 0.05 μm alumina. Subsequently, the samples were annealed in a high-vacuum furnace at 900 °C for 8 hours to relieve strain from rolling and polishing.

The salt mixture was prepared from 99.99% LiF, 99.99% NaF, 99.99% KF, and 99.98% $EuF_3$ powders from Alfa Aeser, part of Thermo Fisher Scientific. The purity of LiF, NaF, and KF is specified on a metals basis, while $EuF_3$ purity is given on a rare earth oxide (REO) basis. All salt preparation was conducted inside a glove box under an argon atmosphere, with oxygen and moisture levels maintained below 1 ppm each. Salt baking and melting were performed using glassy carbon (GC) crucibles from HTW Germany. The individual salt powders (LiF, NaF, and KF) were first baked at 200 - 300°C for more than 12 hours, followed by melting at 900 °C (LiF, KF) and 1000 °C (NaF) for 2 hours. $EuF_3$ was baked at 900 °C for 2 hours. After melting, the salt disks (LiF, NaF, and KF) were crushed into smaller pieces using a stainless-steel mortar and pestle and mixed following a composition of 27.75 wt. % LiF, 11.11 wt. % NaF, 56.14 wt. % KF, and 5 wt. % $EuF_3$. The LiF-NaF-KF ratio follows their eutectic composition, known as FLiNaK, while $EuF_3$ is added to increase the redox potential of the salt, rendering it more corrosive and suitable to study Cr dealloying from Ni-Cr alloys on hour-long timescales[40,41]. $EuF_3$ is expected to corrode Cr, reducing to $EuF_2$ without depositing on the sample surfaces to influence corrosion behavior[42]. The salt mixture was then heated to 700 °C for 12 h to ensure mixing before corrosion testing. Finally, the salt mixture was crushed again using a stainless-steel mortar and pestle to reduce it to smaller pieces suitable for loading into the GC crucibles.

**Corrosion testing**

Corrosion testing was conducted within the same argon-atmosphere glovebox used for salt preparation. Each test utilized an individual GC crucible containing 4.0 g of the salt mixture, sufficient to fully immerse the sample upon melting. The small GC crucible containing the salt and sample was placed into a larger size GC crucible with a cover, which was then nested inside an alumina crucible and loaded into the furnace. The furnace temperature was raised up to the designated corrosion temperature (in the range 700-800°C, depending on sample) and maintained for 8 hours in all tests. After cooling, the sample was extracted from the solidified salt by gently breaking the salt with a stainless-steel mortar and pestle, typically requiring a single tap to expose the sample. The sample was then removed and cleaned multiple times with deionized water.

**Characterization experiments**

Initial characterization was conducted using an FEI FEG-Nova scanning electron microscope (SEM) equipped with a Bruker QUANTAX $e^-$ Flash EBSD detector and a Bruker X Flash energy-dispersive X-ray spectroscopy (EDX) detector. EBSD was conducted at an accelerating voltage of 20 kV and a probe current of 50 nA. Data were collected at a step size of 200 nm and a working distance of 18 mm. EDX data were collected at an accelerating voltage of 15 kV and a probe current of 600 pA. EBSD data were processed and analyzed using the MTEX MATLAB package[43]. SEM images in **Supplementary Fig. 2 and 4** were acquired using a Phenom XL SEM operating in backscattered electron (BSE) mode, also at an accelerating voltage of 15 kV.

Ni-16.6Cr exposed to molten FLiNaK + 5 wt. % $EuF_3$ salt at 800 °C for 8 h was chosen for 3D reconstruction of dealloyed volumes. 3D focused ion beam (FIB)-SEM tomography and 3D EBSD were performed using a Thermo Scientific Helios 5 UXe plasma FIB (PFIB) equipped with Oxford Symmetry S2 EBSD and Oxford Instruments Ultim Max 100 EDX detectors. **Supplementary Fig. 6a-c** show the reconstructed region's position and dimensions.

A polished specimen was mounted on a 36˚ pre-tilt holder SEM stub and was placed specifically to hang over the edge. A target location 150 × 75 µm in size was selected from the cross-section at the edge of the specimen. FIB-induced deposition of tungsten at 12 kV was utilized to prepare a 150 µm × 20 µm protection pad above the volume of interest and also to prepare square pads for the placement of fiducial tracking marks milled into them on the top side for FIB tracking, and beside the cross-section face for SEM tracking, as shown in **Supplementary Fig. 6d**. After establishing a partial tungsten protection layer, three shallow lines were milled along the centerline of the protection pad. These were then filled with FIB-induced deposition of carbon, followed by an additional protection pad deposition of tungsten to approximately 15 µm total height. The carbon lines within the pad cross-section produce an internal high-contrast mark against the tungsten background that is consistently shaped and positioned to facilitate reliable auto focus and auto astigmatism functionality at this location during the SEM image acquisition step. This is critical for fast, accurate, and automated FIB serial sectioning to come.

Following all deposition steps, gaps were milled using the PFIB at 30 kV on the left and right sides of the volume of interest, and also under the target volume as shown in **Supplementary Fig. 6e**. These gaps mitigate shadowing during image acquisition, provide additional space for milling debris to escape, and retain an unobstructed visual pathway towards the EBSD camera. Tomographic acquisition was performed using the Auto Slice and View 5 (ver. 5.9) software by Thermo Fisher Scientific Inc. The automated acquisition was set up for a milling slice interval of 50 nm, across the length of approximately 20 µm, using a 30 kV and 4 nA ion beam. SEM images were acquired for each slice using the concentric backscattered detector, with voltage at 30 kV to optimize for EBSD acquisition, and a 50 nm pixel size to match the slice thickness (i.e., 50 nm voxel size). EBSD maps were also acquired during the process at a spacing of 200 nm between maps, or every 4 slices. The automated acquisition acquires an SEM image angled facing the cross-section plane, then acquires an EBSD map at a 70˚ cross-section face tilt angle with tilt correction applied, and then the stage is moved for FIB milling to remove a 50 nm slice of material from the cross-section face. This automated process is repeated for each slice, except for EBSD maps were acquired every fourth slice. A total of five EDX maps were acquired at the positions of Z=0, 5, 10, 15 and 20 μm; Z direction is marked in **Supplementary Fig. 6b**. The microscope setup for PFIB milling and data acquisition is shown in **Supplementary Fig. 6f**.

**Grain size and twin boundary analysis**

Grain size and twin boundary density were quantified separately in the dealloyed and unaffected regions identified from the cross-sectional micrographs and corresponding EBSD maps. Grain size was estimated by the line-intercept method using five test lines in each region. Only the effective line segments within the solid material were included, while voids, crevices, and unrelated features were excluded. The apparent grain size was calculated as $\bar{d} = \frac{\sum L_i}{\sum N_i}$, where $L_i$ is the effective length of the $i$th test line, $N_i$ is the number of GB intersections along that line, and $i$ indexes the five test lines. Twin boundary density was quantified as $\rho_{\mathrm{TB}} = \frac{N_{\mathrm{TB}}}{A}$, where $N_{\mathrm{TB}}$ is the number of twin boundaries counted within the analyzed region and $A$ is the corresponding analyzed area. Boundary length was not used for this quantification because the twin boundaries were frequently curved and irregular, making length measurements less reliable.

**3D correlated FIB–SEM tomography and EBSD**

**Supplementary Fig. 7** summarizes the 3D reconstruction and analysis pipeline. SEM images acquired from large-volume 3D FIB-SEM tomography often show curtaining effects, especially with porous materials, so we suppressed these stripe-like artefacts using Fourier-domain filtering (**Supplementary Fig. 8**) by masking the curtaining-related frequency bands in the 2D FFT and reconstructing the filtered images via inverse FFT. The filtered stack was then corrected for slice-to-slice misalignment (**Supplementary Fig. 9**) using intensity-correlation-based registration to remove drift, followed by correction of a systematic y-direction shift by aligning the top boundary of the cross-sectional region across the stack. The images were then cropped to the region of interest.

Void space was segmented from the BSE stack using nnU-Net (**Supplementary Fig. 10**). A subset of slices was manually labeled to train the model, which was then applied to the full stack; segmentation quality was verified by overlaying predicted void contours onto the original BSE images. EBSD maps were correlated to the BSE volume by registering each EBSD slice to its corresponding BSE slice using voids as fiducial markers (**Supplementary Fig. 11**). Voids were extracted from EBSD by thresholding, and in-plane scaling plus translation were optimized by maximizing void-pixel overlap, after which the same transformation was applied to the full EBSD maps and validated by overlay.

GB networks were segmented from the registered EBSD series using a separate nnU-Net model (**Supplementary Fig. 12**) trained on manually annotated GB masks. Accuracy was assessed by overlaying extracted GB lines onto the original EBSD maps. Twin boundaries were identified by crystallographic twin analysis and removed from the GB-line images by subtraction to yield general GB networks (**Supplementary Fig. 13**). The resulting GB lines were denoised by connected-component area filtering to remove small, isolated artefacts (**Supplementary Fig. 14**), then thickened by morphological dilation for 3D visualization (**Supplementary Fig. 15a–d**). To reduce discontinuities caused by the larger spacing between EBSD sections, intermediate GB frames were inserted using signed-distance-transform-based interpolation (**Supplementary Fig. 15e–g**). The 3D GB volume was reconstructed by stacking the processed GB images (**Supplementary Fig. 16**) and further denoised using connected-component volume filtering to remove small, isolated volumes.

Void structures were reconstructed in 3D from the segmented BSE images (**Supplementary Fig. 17a**). To remove oxides and precipitates misclassified as voids, we applied feature-level filtering based on specific surface area (**Supplementary Fig. 17b**), removing thin oxide-like features with abnormally high specific surface area and precipitate-like features with comparatively low specific surface area. Connectivity between voids and GBs was quantified in Dragonfly by connected-component analysis on a combined binary volume (**Supplementary Fig. 18**). Voids in the same connected component as the GB network were classified as GB-connected, and all others as non-GB-connected. Finally, residual artefacts not removed by automated filtering were manually cleaned by visual inspection (**Supplementary Fig. 19**) to obtain the final void dataset.

**Author contributions**

W.Z., H.G., Y.Y., M.P.S., and S.Y.P. conceived of the project, while K.D., M.J., and M.R.D. provided critical guidance on the project. W.Z. performed the corrosion testing and initial SEM characterization. H.G. conducted additional SEM and EBSD characterization experiments. H.G., T.C., and M.M. performed the 3D FIB-SEM tomography experiments. L.D., Y.Z., and Y.Y. performed the correlated FIB–SEM tomography and EBSD reconstruction, conducted the quantitative data analysis, and designed and prepared Figs. 1–5 and Supplementary Figs. 5 and 7–19. W.Z. and H.G. wrote the original draft of the manuscript. Y.Y. substantially rewrote and reorganized the manuscript. All authors contributed to the reviewing and editing of the manuscript.

**Acknowledgements**

The authors would like to acknowledge the funding from Eni S.P.A., Italy through the MIT Plasma Science and Fusion Center's Laboratory for Innovative Fusion Technologies (LIFT). In addition, this work was funded in part by the University Network of Excellence in Nuclear Engineering (UNENE) Research Chair in Corrosion Control and Materials Performance, the Canadian Nuclear Safety Commission (CNSC) (Grant ALLRP 580446-2022), and the Natural Sciences and Engineering Research Council of Canada (NSERC) Alliance Program (Grant ALLRP 544898-19). L.D., Y. Z., and Y.Y. were supported by the National Science Foundation (NSF) CAREER award DMR-2145455. The authors thank the staff at the Reactor Materials Testing Laboratory (RMTL) at Queen's University for providing access to microscopy facilities and important assistance.

**Data availability**

All original SEM/EDX images and EBSD data, including those acquired from 3D FIB-SEM tomography, necessary to reproduce the results presented in this manuscript, are available in the open data repository associated with this work (https://doi.org/10.5281/zenodo.15546952). Due to the large file size, the reconstructed 3D tomography datasets are available from the corresponding authors upon reasonable request.

## Competing interests

The authors have no competing interests to declare, financial or otherwise, which could have affected the quality or outcome of this work.

## Statement on the use of generative AI

During the preparation of this work, a subset of the authors used ChatGPT in order to improve grammar/language and clarity of expression. After using ChatGPT, the authors reviewed and edited the content as needed, and all authors take full responsibility for the content of the published article.

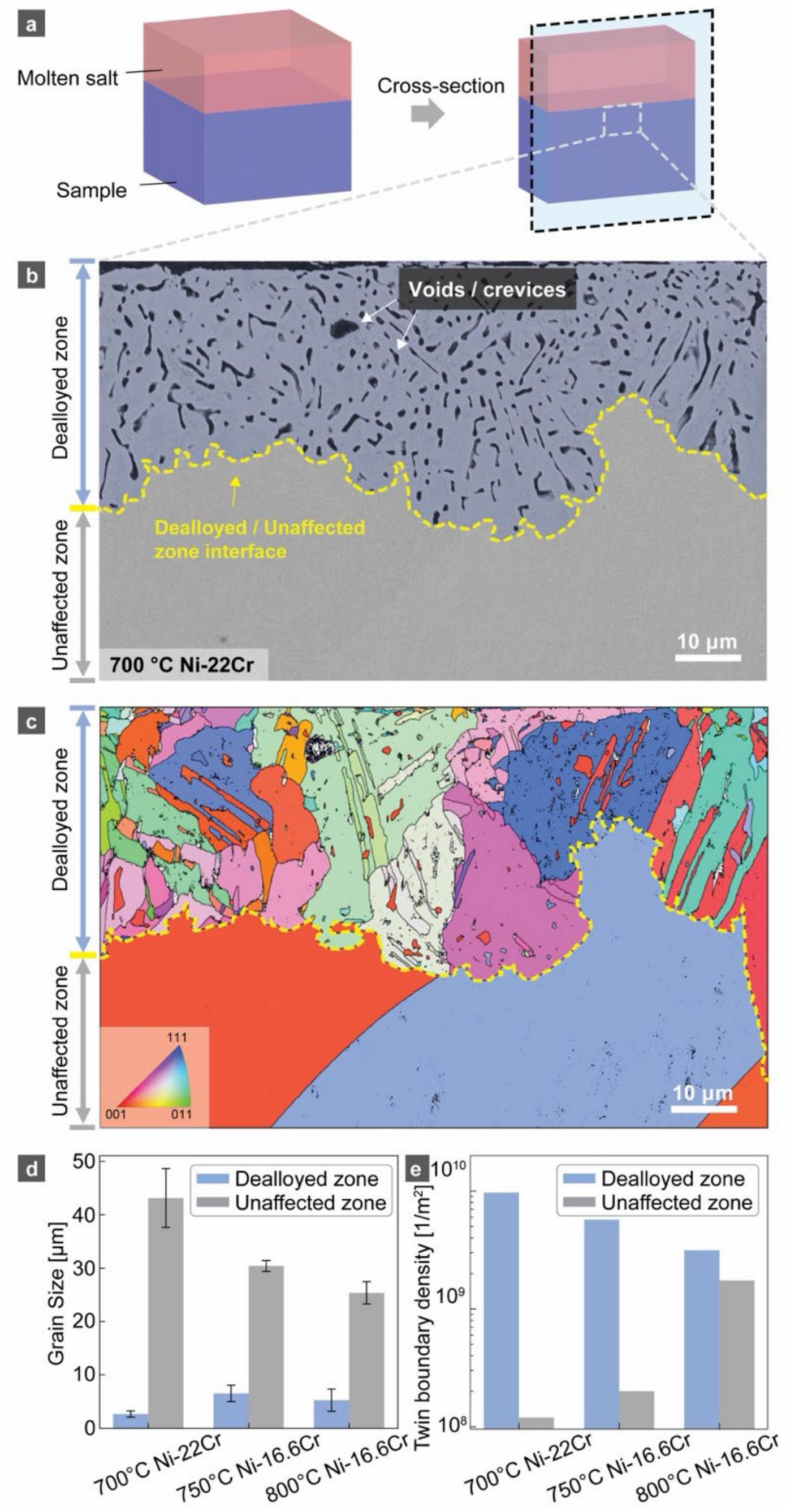

**Fig. 1 | Correlated SEM and EBSD reveal dealloying-induced recrystallization. a,** Schematic illustrating the cross-sectioning geometry and the region selected for SEM/EBSD analysis after molten-salt exposure. **b,** Cross-sectional BSE-SEM image showing the corroded (dealloyed) region, identified by a high density of voids and crevices relative to the underlying, unaffected alloy. **c,** EBSD map of the same region as in **b**, showing pronounced grain refinement within the corroded zone and crystallographic orientations distinct from the parent grains beneath the dealloyed layer. **d,** Grain-size statistics comparing dealloyed and unaffected regions across the three tested conditions, demonstrating systematic grain refinement after dealloying. **e,** Quantitative twin-boundary (Σ3) density (or fraction) in dealloyed versus unaffected regions for the same three conditions, showing enhanced Σ3 content associated with the refined-grain dealloyed zone.

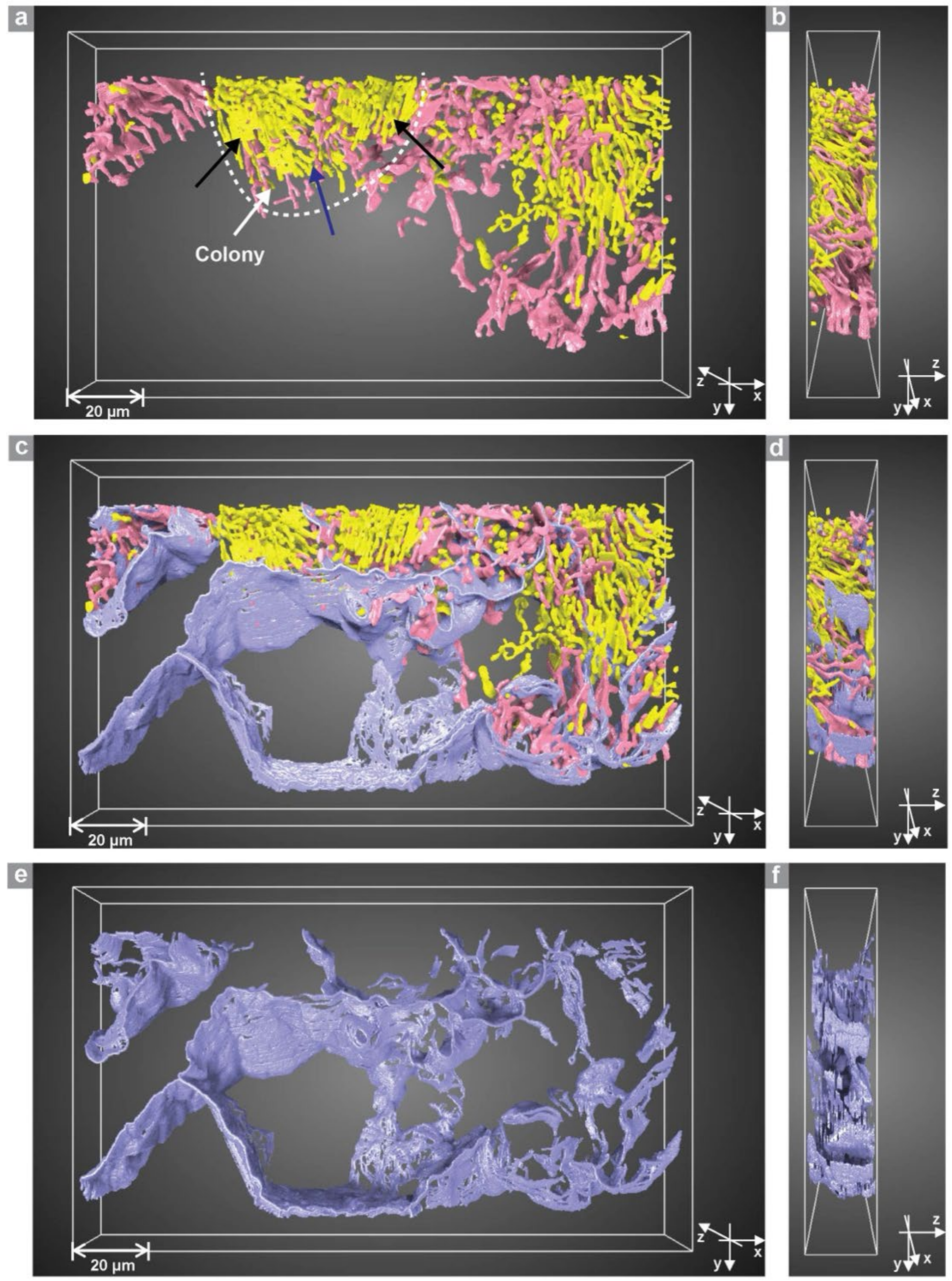

**Fig. 2 | 3D GB–void architecture reconstructed by correlated FIB–SEM tomography and EBSD. a, b,** 3D rendering of the reconstructed 3D void space from FIB–SEM serial sectioning, highlighting the predominantly interconnected porosity that supports salt infiltration. White dashed outline in **a** highlights representative dealloying colony showing the overall half-ellipsoidal morphology and the tunnel-like void network near the GB-guided corrosion front. **c, d,** Co-registered 3D crystallographic reconstruction from EBSD, showing grains and the corresponding GB network within the same volume as in **a** and **b**. **e, f**, 3D crystallographic reconstruction from EBSD, showing the GB network. *(Supplementary Videos show the same volume renderings and connectivity in 3D.)*

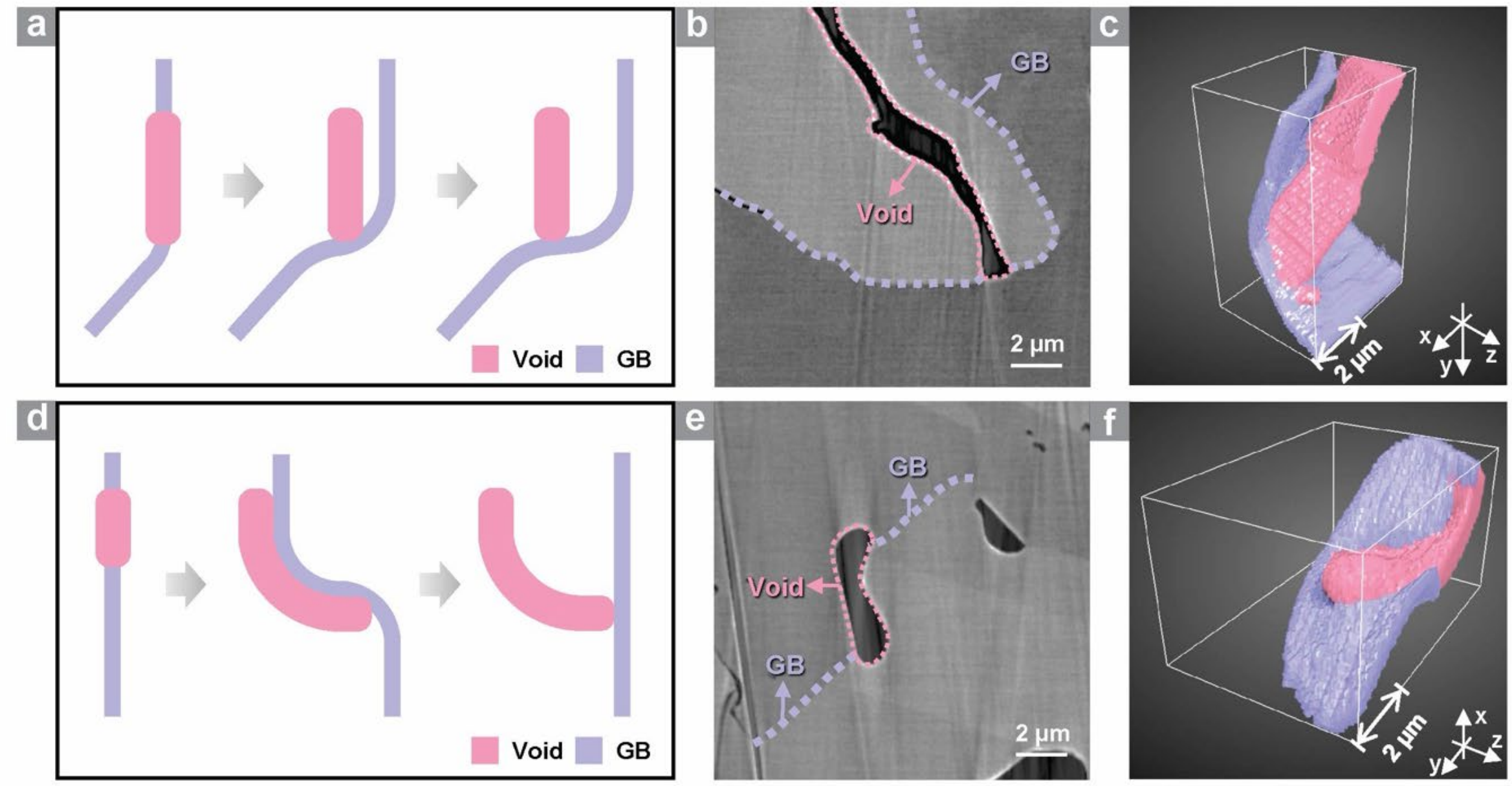

**Fig. 3 | Reorientation pathways from GB-parallel wormholes to GB-perpendicular tunnels. a,** Schematic of a "GB shake-off" pathway: a tunnel that initially lies along and parallel to a GB becomes embedded within a grain as the GB migrates away, leaving a remnant tunnel that remains parallel to the former boundary trajectory. **b,** Cross-sectional SEM image showing a detached, GB-parallel tunnel consistent with the shake-off mechanism in **a**. **c,** 3D reconstruction confirming that the tunnel no longer coincides with the active GB plane but preserves a GB-parallel geometry and remains connected within the pore network. **d,** Schematic of an "L"-shaped reorientation pathway driven by differential migration of adjacent GB segments: unequal migration rates bend an initially GB-parallel tunnel into an L shape, followed by detachment that yields a tunnel tip effectively perpendicular to the GB. **e,** Cross-sectional SEM image showing an L-shaped tunnel morphology. **f,** 3D reconstruction validating the bent-and-detached geometry in 3D and its relationship to the local GB configuration.

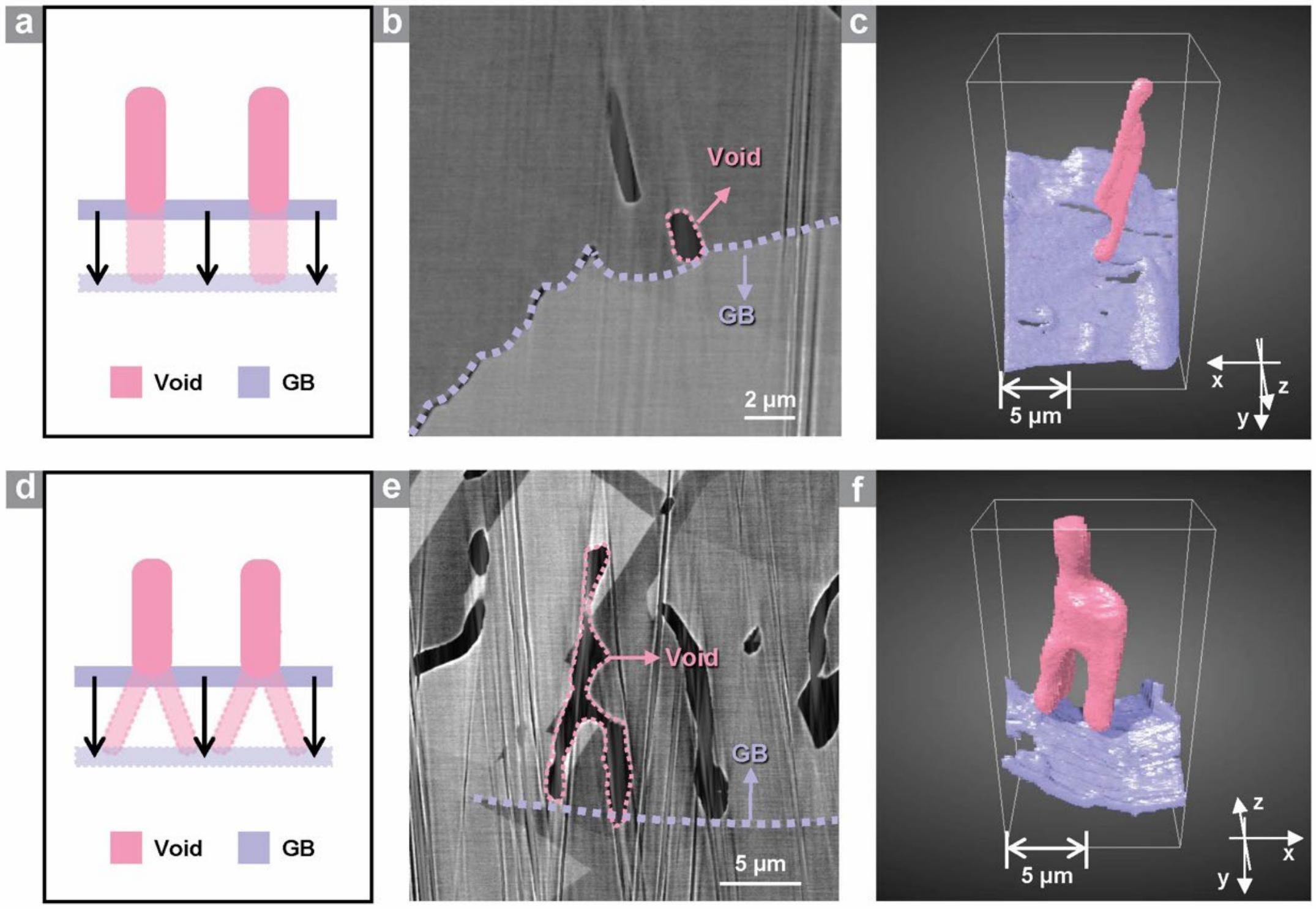

**Fig. 4 | Front-coupled tunnel elongation and branching produce discontinuous precipitation (DP) -like morphologies. a,** Schematic illustrating tunnel elongation when a salt-filled tunnel is oriented approximately perpendicular to the migrating GB: forward GB motion drags the tunnel tip, extending the tunnel behind the front. **b,** Cross-sectional SEM image showing elongated, front-perpendicular tunnels consistent with the mechanism in **a**. **c,** 3D reconstruction of the same morphology, directly visualizing tunnel elongation and connectivity within the colony interior. **d,** Schematic illustrating branching during front-coupled growth, in which a perpendicular tunnel splits into multiple branches as the GB advances. **e,** Cross-sectional SEM image capturing a representative branching event. **f,** 3D reconstruction confirming branching in 3D and illustrating the resulting DP-like array and connectivity of tunnels.

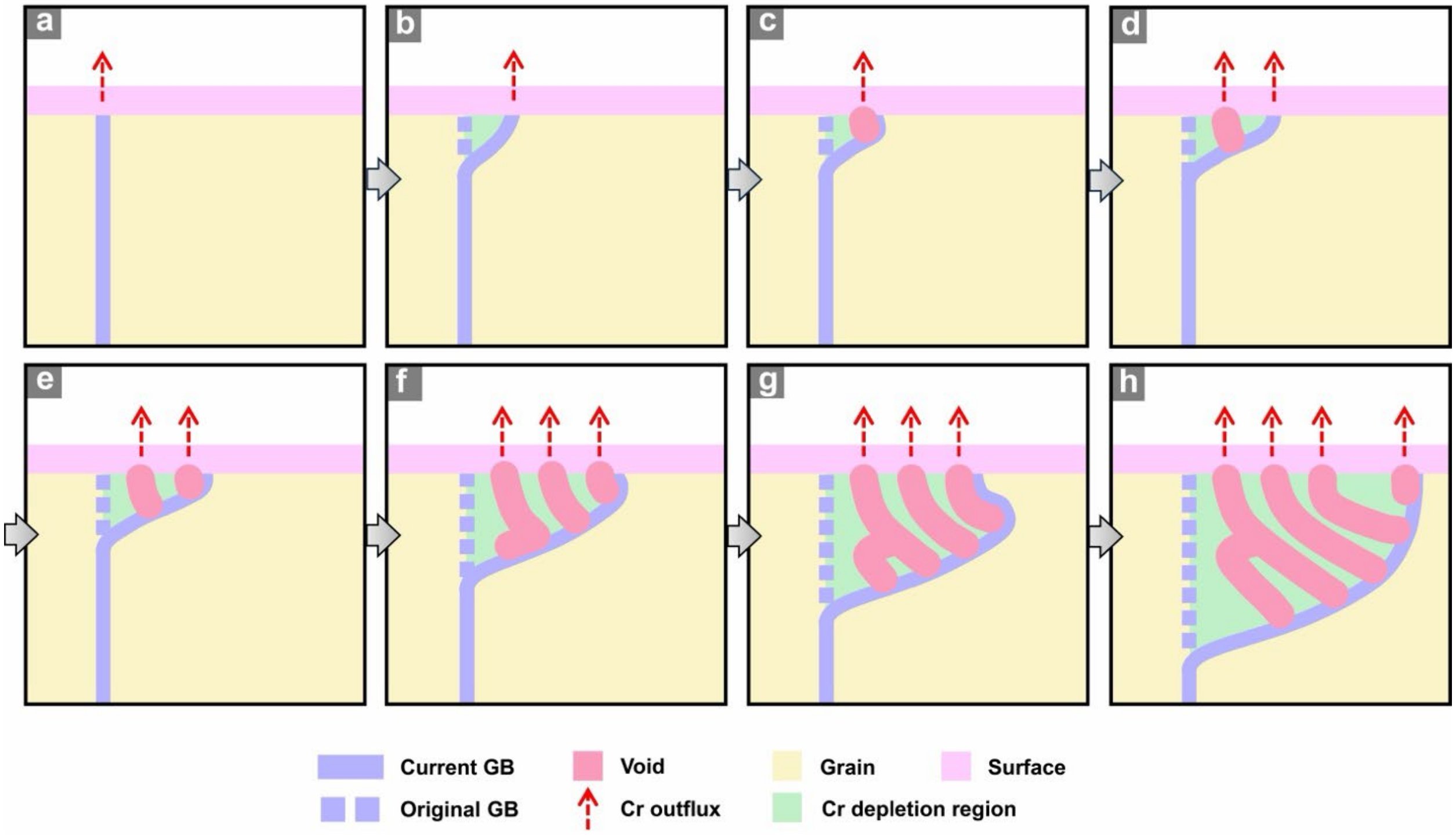

**Fig. 5 | Cross-sectional schematic showing proposed GB-assisted bulk dealloying mechanism in molten salt. a,** preferential Cr removal along pre-existing GBs; **b,** DIR/DIGM-driven GB migration and refinement; **c,** surface-initiated wormholes growing along GBs and filling with salt; **d,** tunnel detachment by "GB shake-off"; **e,** repeated wormhole formation and shake-off during continued GB migration; **f,** DP-like elongation of GB-perpendicular tunnels as the front advances; **g,** occasional branching during tunnel growth; **h,** fully developed interconnected tunnel–GB network that sustains bulk dealloying below the parting limit.

# Supplementary Information

## Anomalous bulk dealloying below the parting limit

Weiyue Zhou[1#,*], Hooman Gholamzadeh[2#], Lei Ding[3#], Kevin Daub[2], Mehdi Mosayebi[4], Travis Casagrande[4], Yingxin Zhu[3], Miaomiao Jin[5], Mark R. Daymond[2], Yang Yang[3,*], Michael P. Short[1,*], Suraj Y. Persaud[2,6,*]

1. *Department of Nuclear Science and Engineering, Massachusetts Institute of Technology, 77 Massachusetts Avenue, Cambridge, MA 02139, USA*
2. *Department of Mechanical and Materials Engineering, Smith Engineering, Queen's University, 60 Union St. W, Kingston, Ontario K7L 2N8, Canada*
3. *Department of Engineering Science and Mechanics, The Pennsylvania State University, University Park, PA 16802, USA*
4. *Canadian Centre for Electron Microscopy, McMaster University, 1280 Main St. W, Hamilton, Ontario L8S 4M1, Canada*
5. *Department of Nuclear Engineering, The Pennsylvania State University, University Park, PA 16802, USA*
6. *Department of Materials Science and Engineering, McMaster University, 1280 Main St. W, Hamilton, Ontario L8S 4M1, Canada*

* Email addresses of corresponding authors: wyzhou@mit.edu, yangyang@alum.mit.edu, hereiam@mit.edu, persas39@mcmaster.ca.

# These authors contributed equally.

## Table of contents

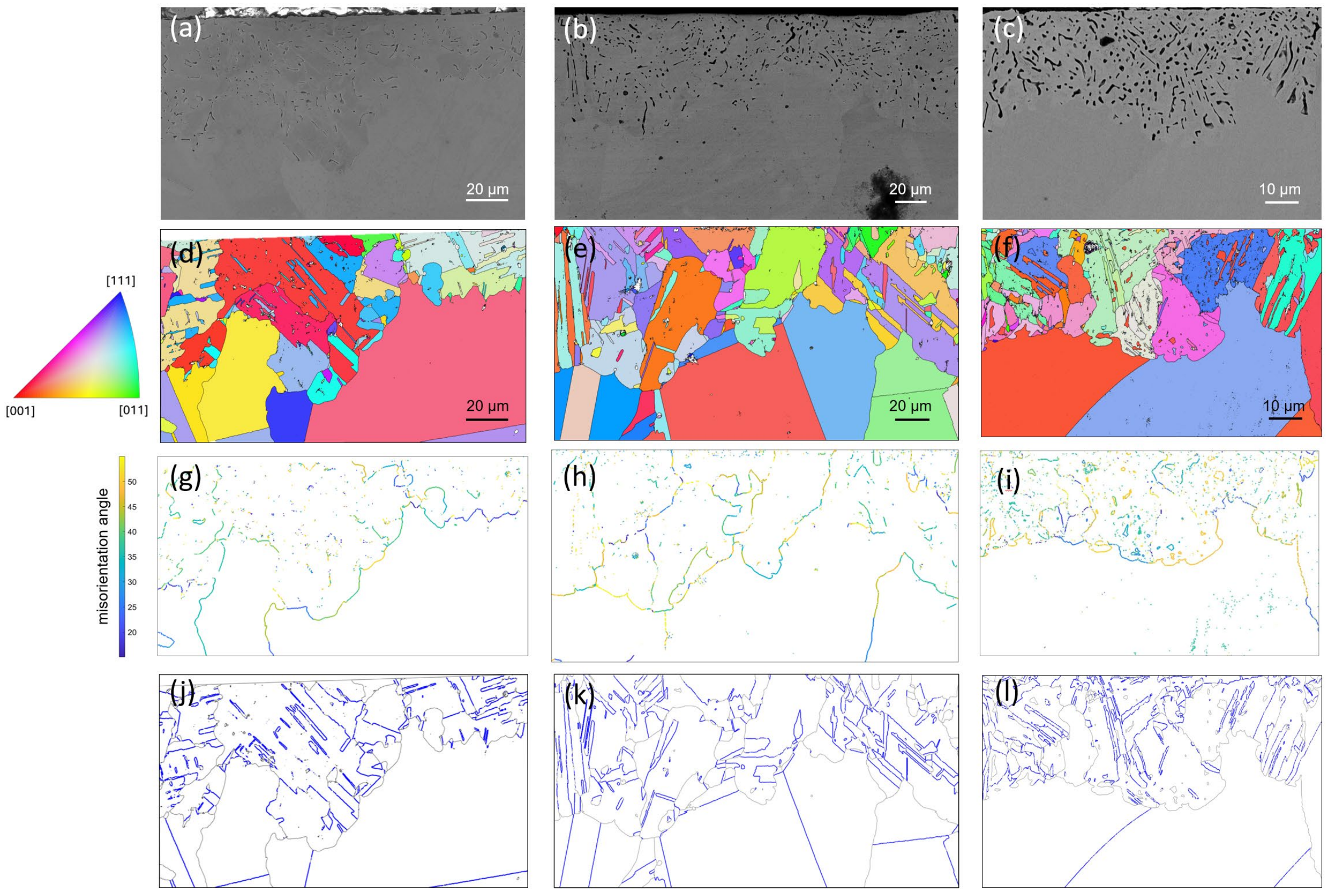

**Supplementary Fig. 1 |** SEM-BSE images and corresponding EBSD images of the cross-sections of NiCr alloys after 8 h exposure to molten fluoride salt. (a),(b) Ni-16.6Cr at 750 and 800 °C, respectively.(c) Ni-22 Cr at 700 °C. (d)-(f) IPFZ maps, (g)-(h) High angle GBs (15° - 55°), excluding Σ3 GBs, and (j)-(l) Σ3 GBs corresponding to the BSE images in (a)-(c) respectively.

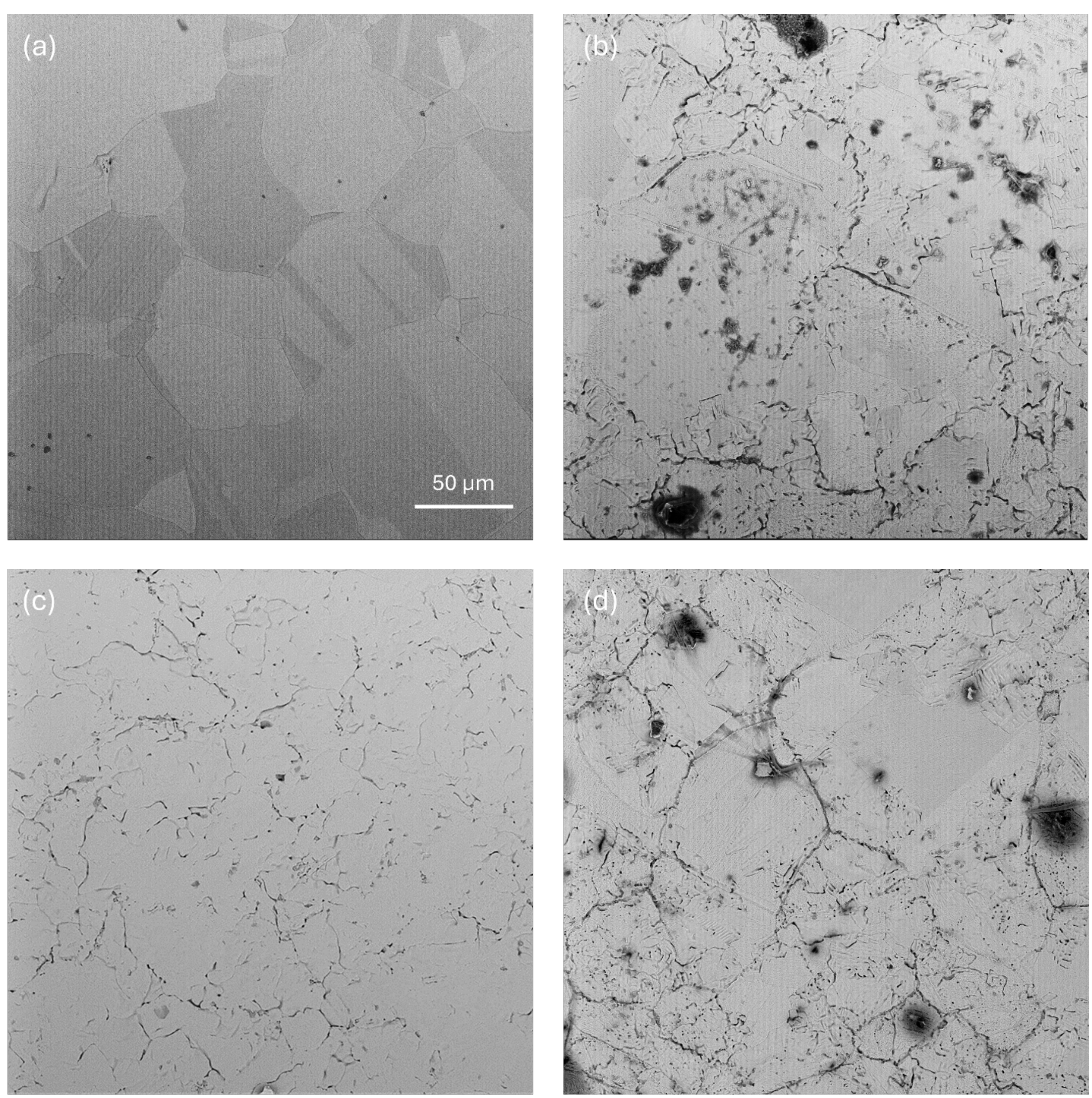

**Supplementary Fig. 2** | SEM-BSE images from Ni-Cr alloy surfaces before and after 8 h molten salt exposure. (a) The surface of a pristine Ni-16.6 Cr alloy. (b) The surface of a Ni-16.6 Cr alloy after exposure at 750 °C. (c) The surface of a Ni-16.6 Cr alloy after exposure at 800 °C. (d) The surface of a Ni-22Cr alloy after exposure at 700 °C.

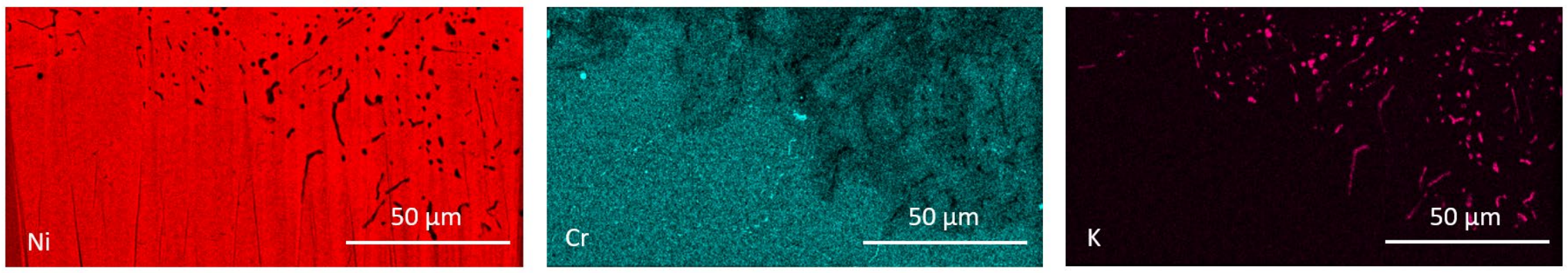

**Supplementary Fig. 3** | SEM-EDX elemental maps of Ni and K for the Ni-16.6 Cr alloy cross-section after exposure at 800 °C. The presence of K highlights salt infiltration along the crevices.

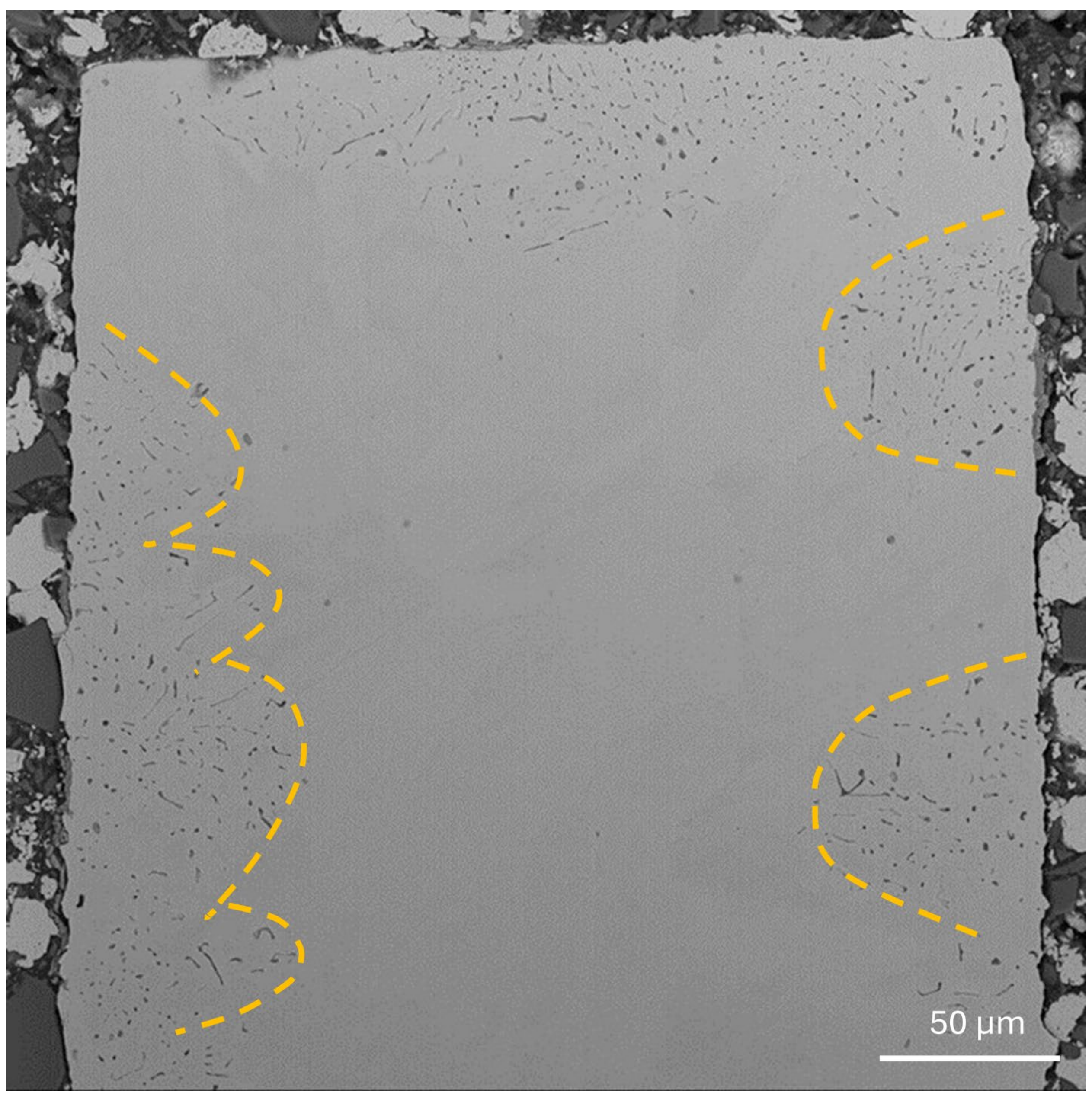

**Supplementary Fig. 4** | SEM-BSE image from the cross-section of a Ni-16.6 Cr alloy after 8 h exposure to molten fluoride salt at 800 °C showing half ellipse colonies of the dealloying/tunnels.

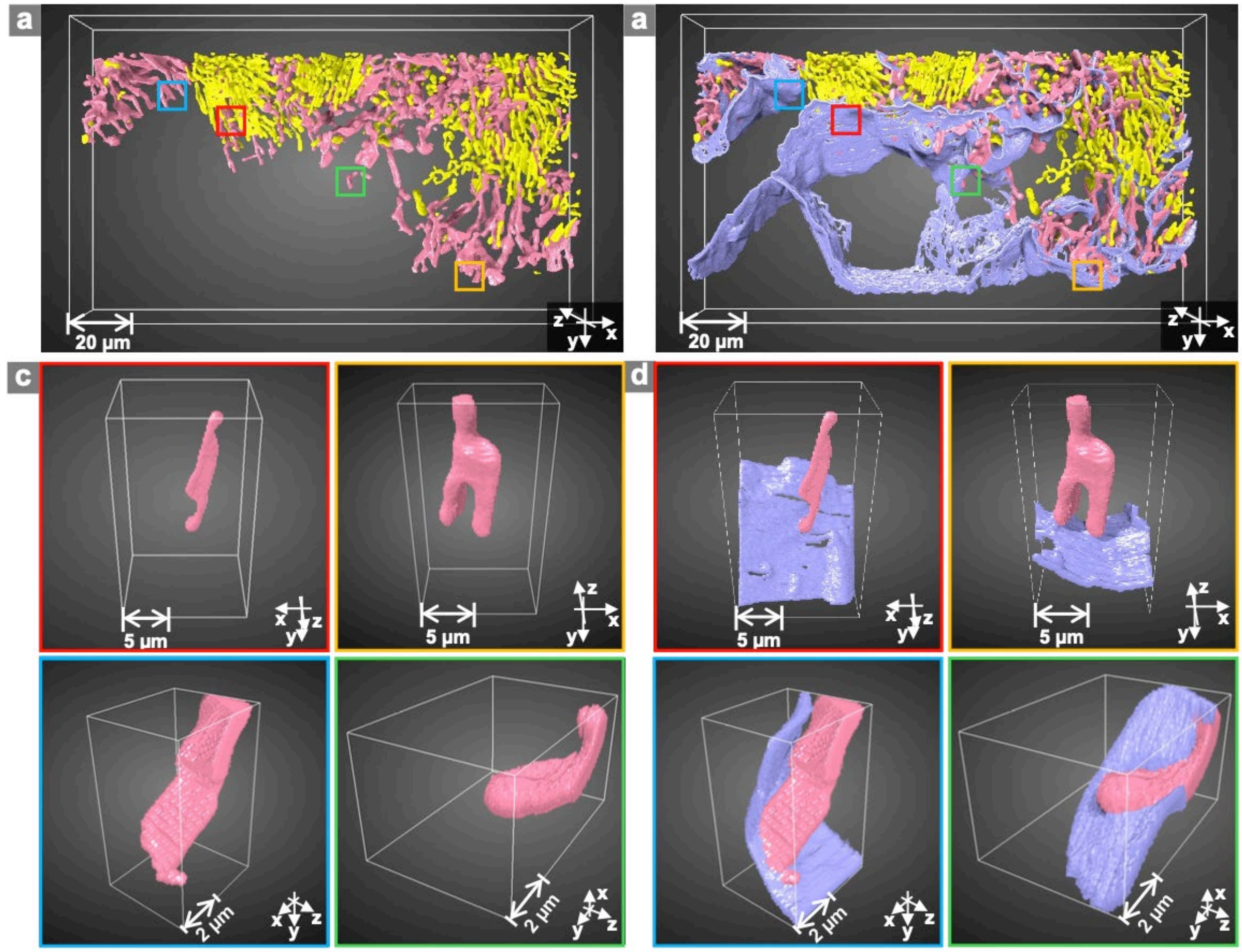

**Supplementary Fig. 5** | Overview of void and GB 3D structures and locations of selected voids for illustration in Figs. 3-4. (a) Overview of void 3D structures, with red color indicating void structures connected to GBs and yellow color indicating void structures not connected to GBs. (b) Overview of void and GB 3D structures, with red color indicating void structures connected to GBs and yellow color indicating void structures not connected to GBs. (c) Selected local voids for illustration corresponding to the locations marked by red, orange, blue, and green squares in (a). (d) Selected local voids and associated GB for illustration corresponding to the locations marked by red, orange, blue, and green squares in (b).

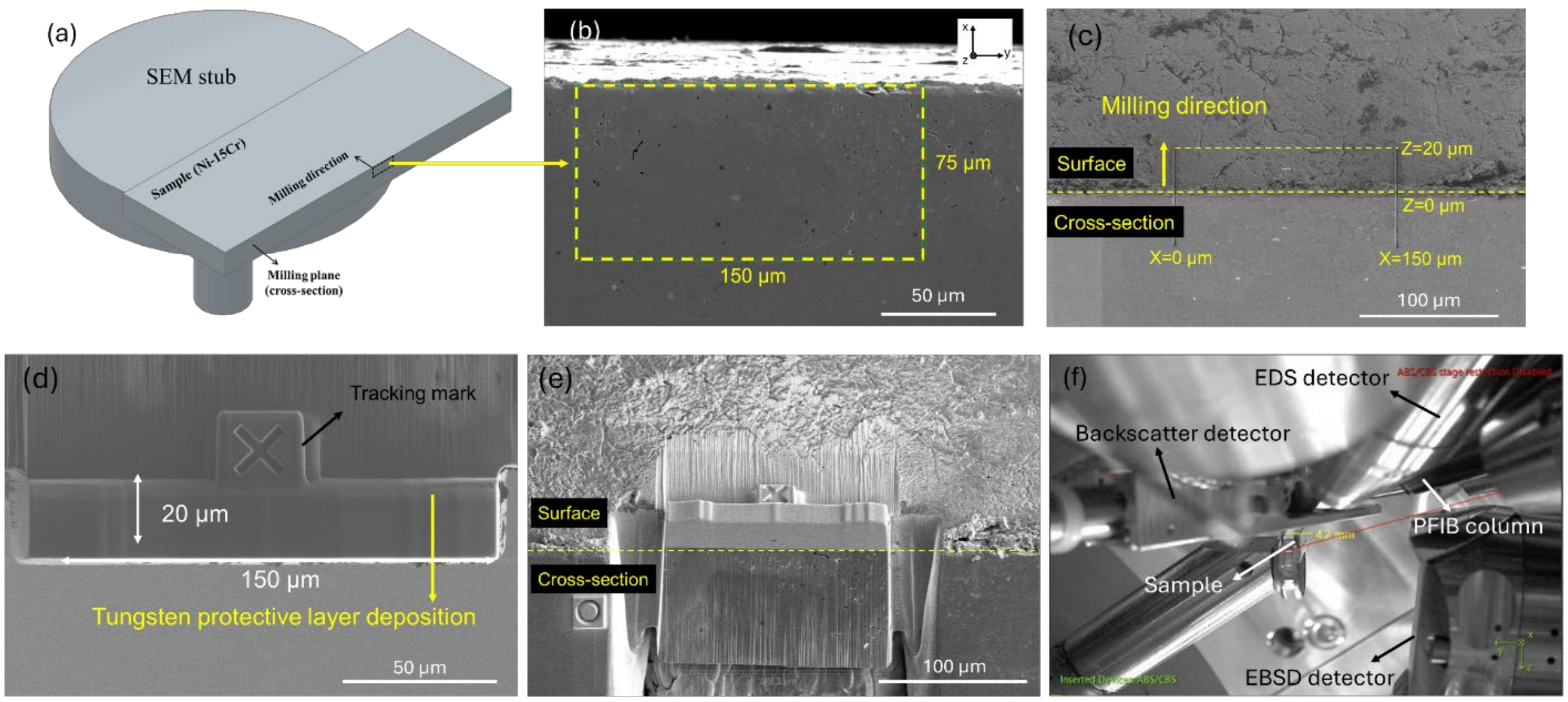

**Supplementary Fig. 6** | PFIB-SEM processes. (a) Sample set-up on the stub for PFIB milling. (b) Position and dimensions of ROI for PFIB milling. (c) Direction of PFIB milling. (d) Tungsten coating on the top to protect the ROI for PFIB milling. (e) Gaps milled on the side and under the ROI. (f) Microscope set-up for PFIB milling and data acquisition.

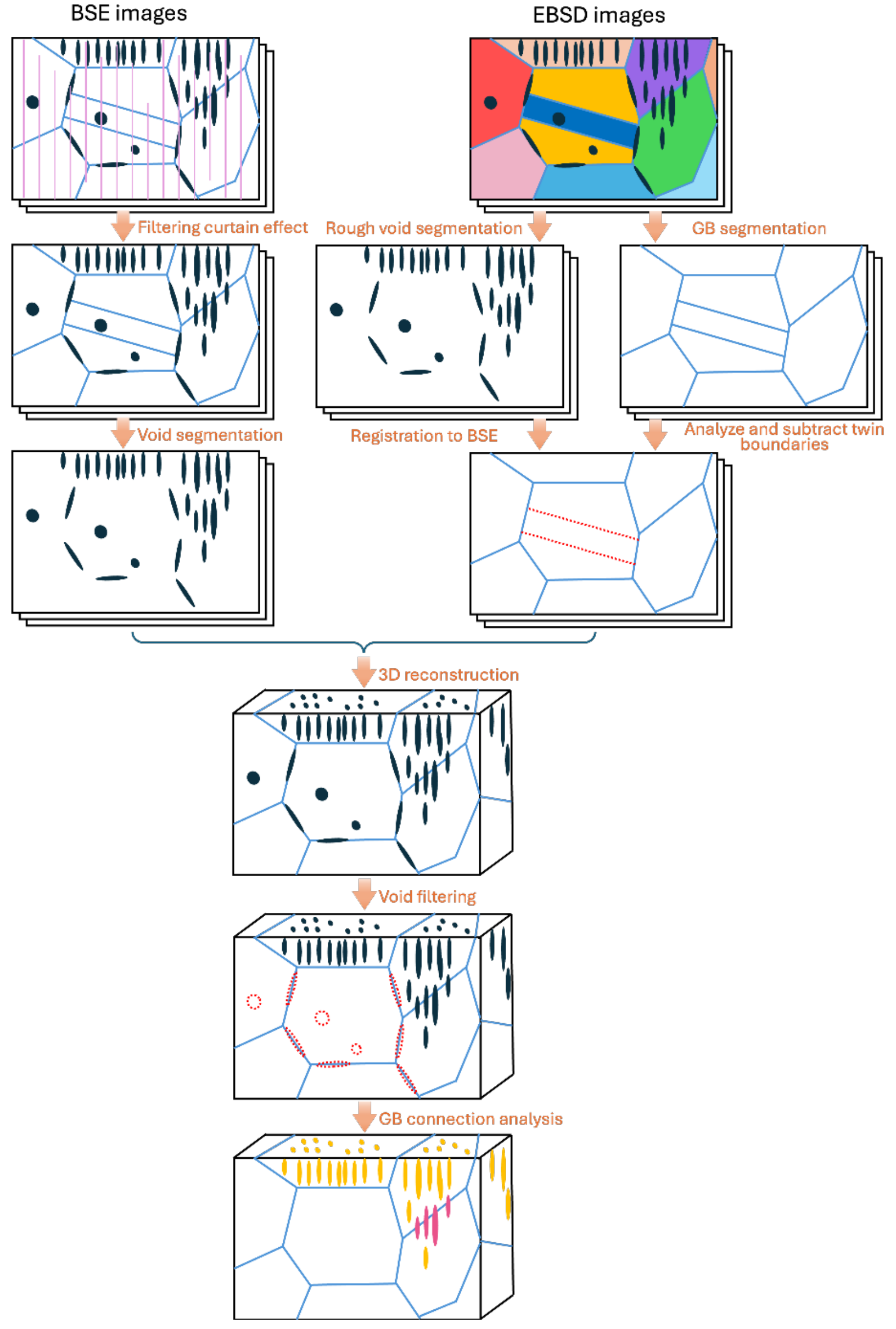

**Supplementary Fig. 7** | Schematic diagram of the 3D reconstruction method.

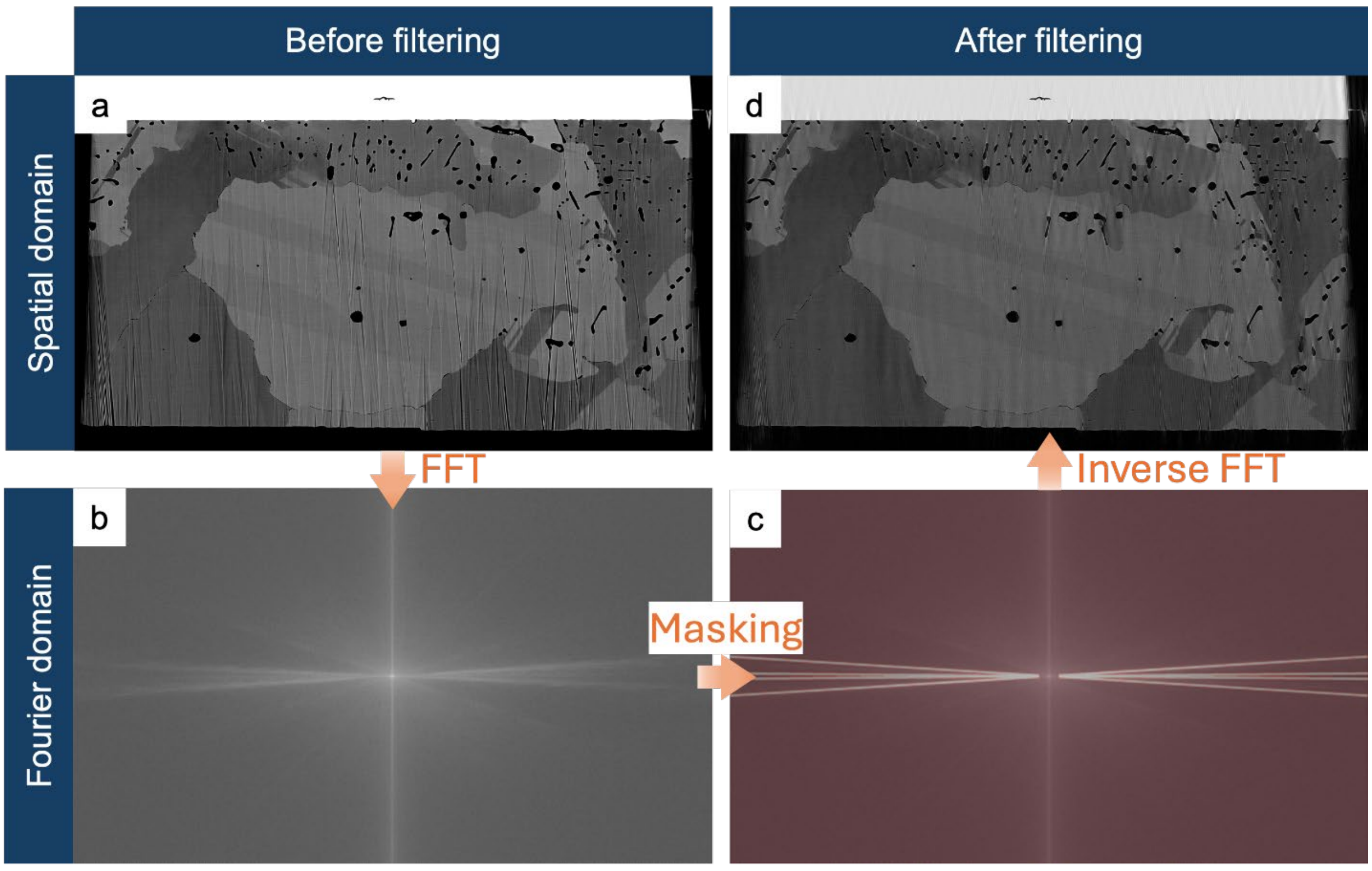

**Supplementary Fig. 8** | Suppression of curtain effect artifacts in BSE images using Fourier-domain filtering. Curtain effect artifacts caused by FIB milling in the BSE images were suppressed by applying a Fourier-domain masking filter. (a) A representative BSE image containing curtain effect stripes. (b) Fourier transform of (a). (c) Specific frequency bands corresponding to the curtain effect stripes were selectively removed in the Fourier domain, followed by (d) inverse Fourier transformation to reconstruct the filtered image with suppressed curtain effect artifacts.

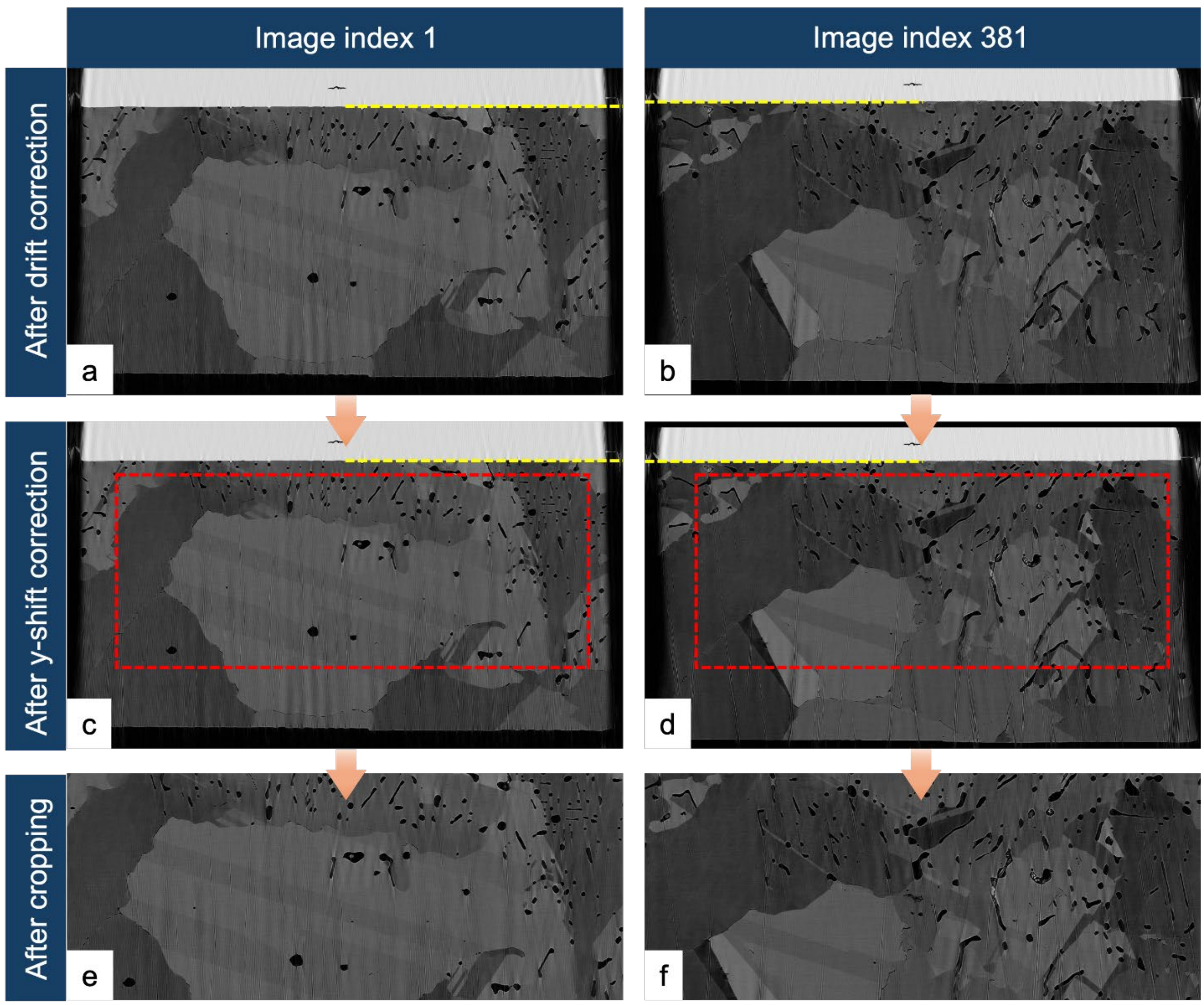

**Supplementary Fig. 9** | Drift correction, y-shift correction, and cropping of the BSE image series. (a) First BSE image (index 1) after drift correction. (b) Last BSE image (index 381) after drift correction. (c) First BSE image (index 1) after y-shift correction. (d) Last BSE image (index 381) after y-shift correction. The yellow lines in images indicate the top boundary of the cross-sectional region. (e) First BSE image (index 1) after cropping. (f) Last BSE image (index 381) after cropping.

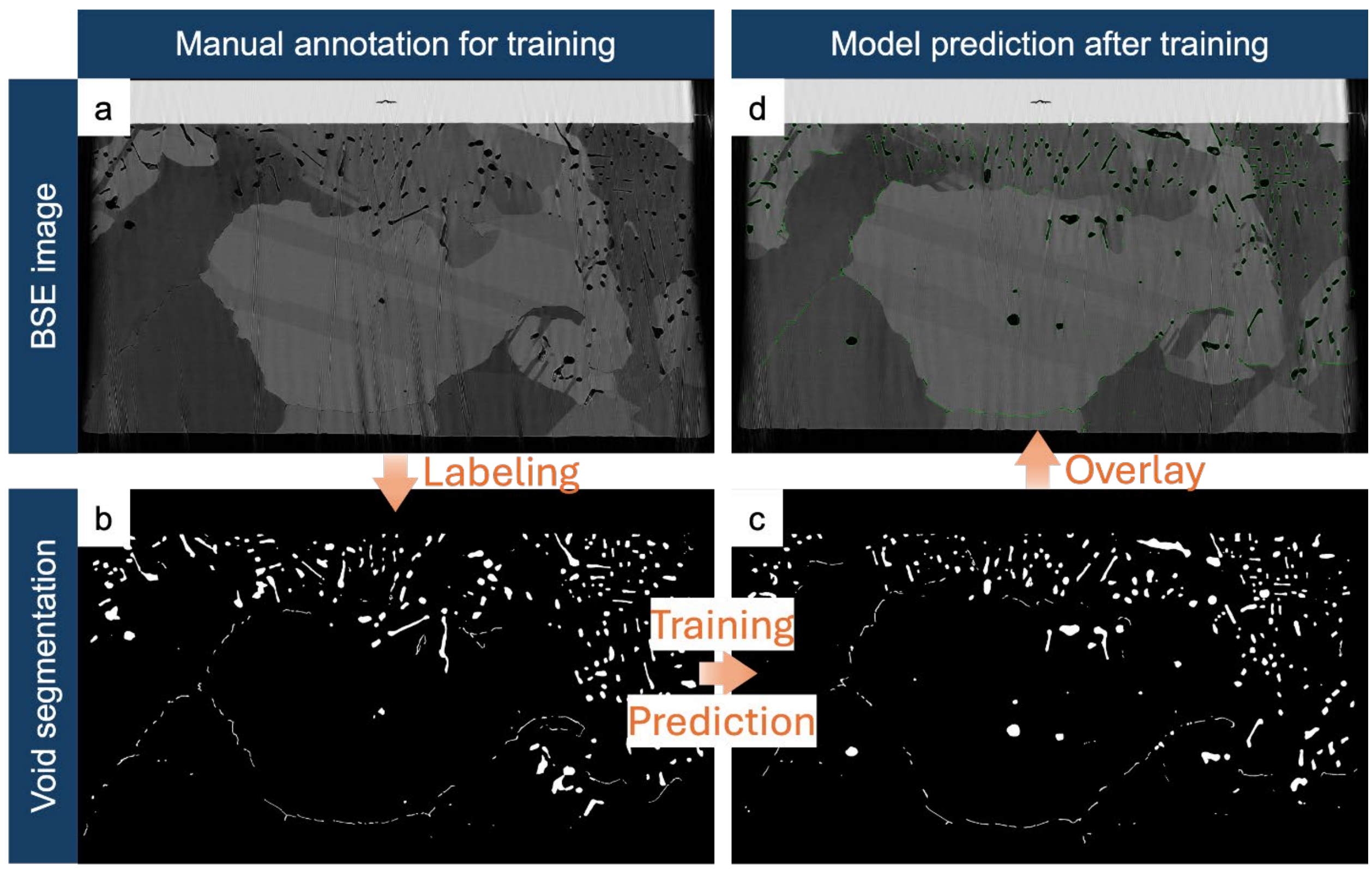

**Supplementary Fig. 10 |** Void segmentation of BSE images using nnUNet. (a) Original BSE image (index 100) used for nnU-Net training. (b) Manual labeling of voids in the same BSE image for training. (c) Example of void segmentation in BSE image (index 9) using the trained nnU-Net model. (d) Overlay of segmented void contours on the original BSE image (index 9).

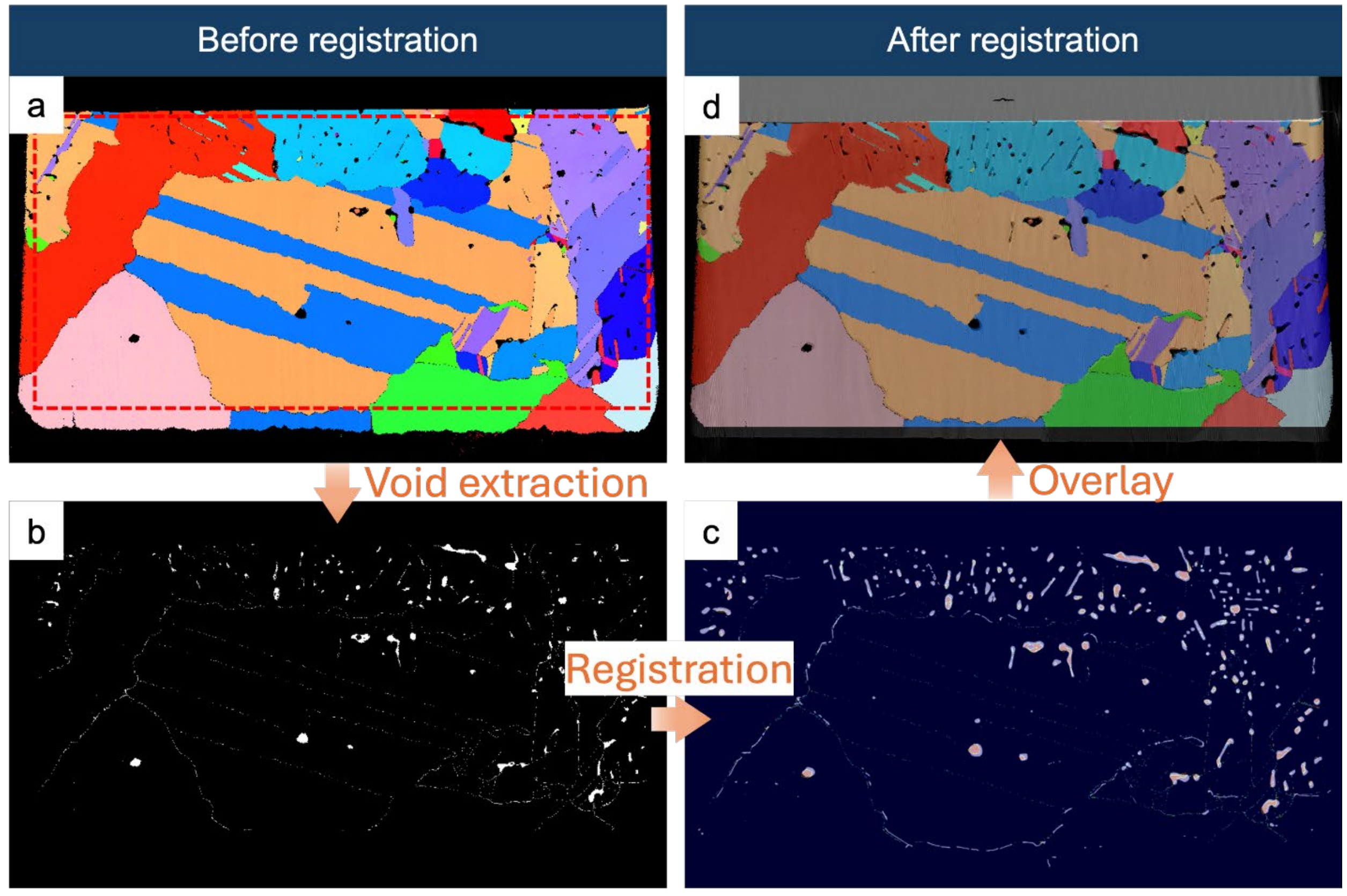

**Supplementary Fig. 11 |** Void-based registration of EBSD images to reference BSE image. (a) Original EBSD image (index 9). (b) Extract voids from EBSD image (index 9) by cropping and setting a brightness threshold. The cropped region is indicated by the red dashed box. (c) Auto-scaling of width and height, along with shift correction, was performed using voids extracted from the EBSD and BSE images (index 9). The transformation parameters were optimized to maximize the overlap between void pixels in the two images. The transformed EBSD voids image is overlaid on the BSE voids image for visual verification. (d) The transformed EBSD image overlaid on the BSE image.

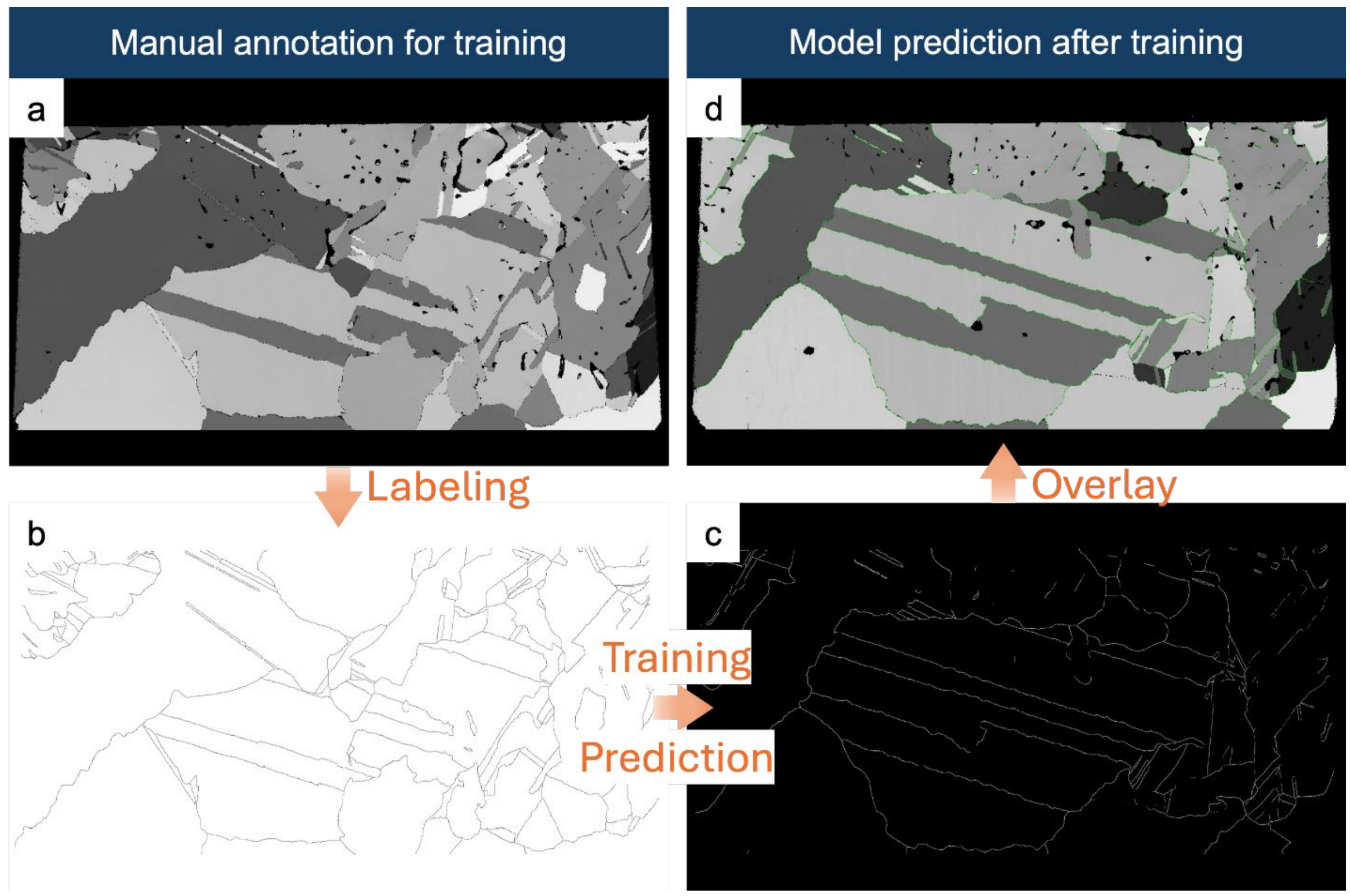

**Supplementary Fig. 12 |** GB segmentation of EBSD images using nnUNet. (a) Original EBSD image (index 301) used for nnU-Net training. (b) Manual labeling of GBs in the same EBSD image for training. (c) Example of GB segmentation in EBSD image (index 9) using the trained nnU-Net model. (d) Overlay of segmented GB lines on the original EBSD image (index 9).

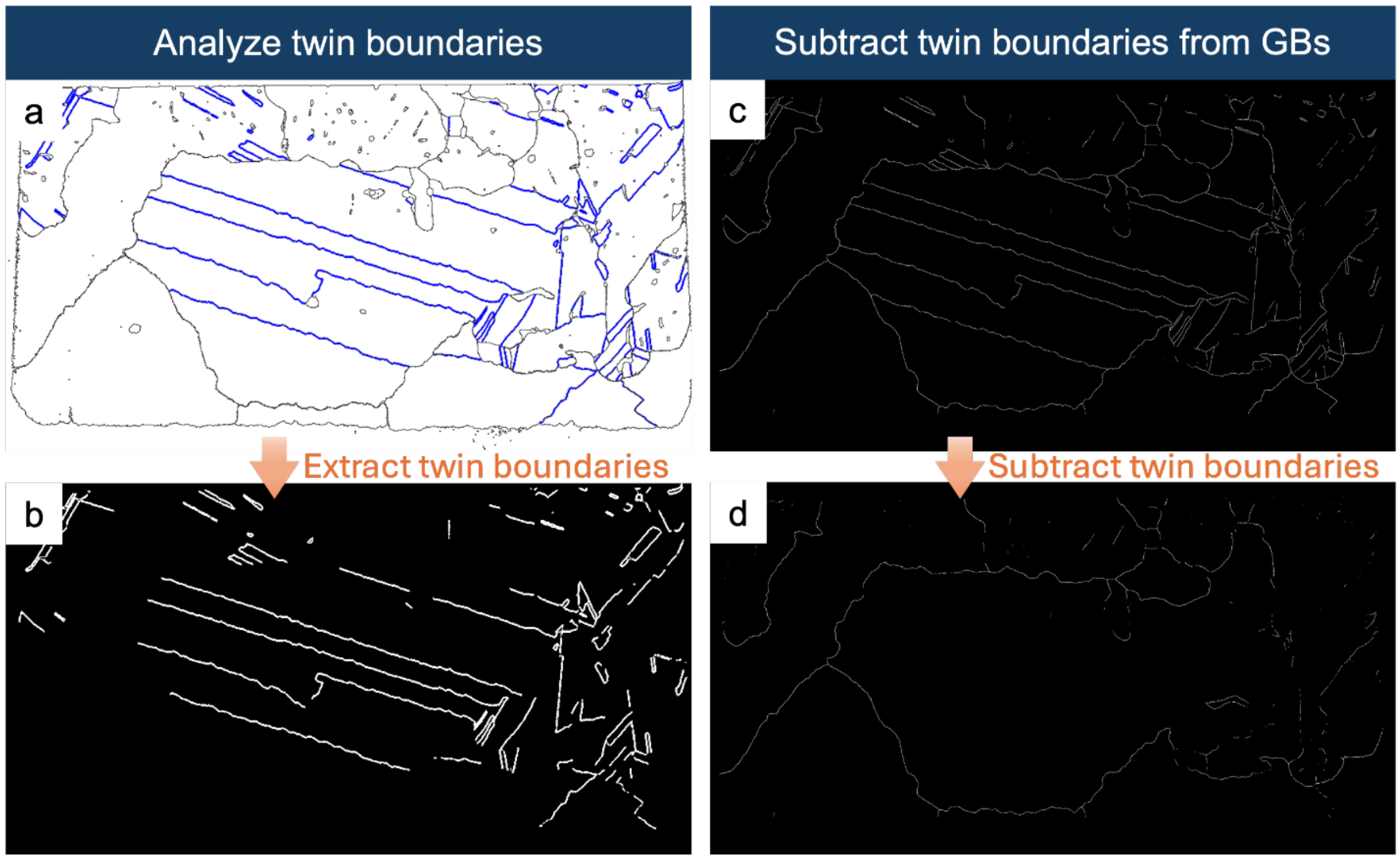

**Supplementary Fig. 13 |** Removal of twin boundaries from GB line image. (a) Twin boundary analysis result of a EBSD image (index 9). (b) Twin boundary lines extracted from (a). (c) GB line image (index 9). (d) GB line image after subtracting the twin boundaries shown in (b) from (c).

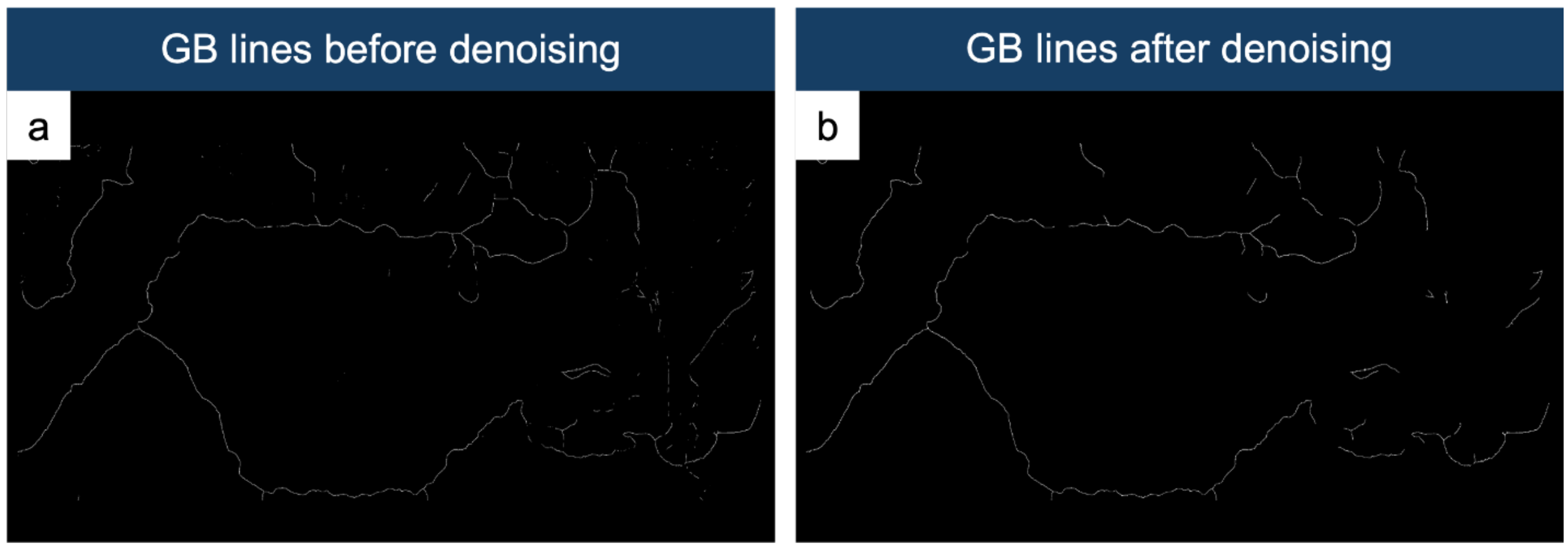

**Supplementary Fig. 14** | Denoising of GB line images. (a) GB line image after removing the twin boundaries (index 9). (b) Denoised image of (a). Noise was removed using a connected-component–based area filtering approach, in which small isolated regions below a threshold were eliminated.

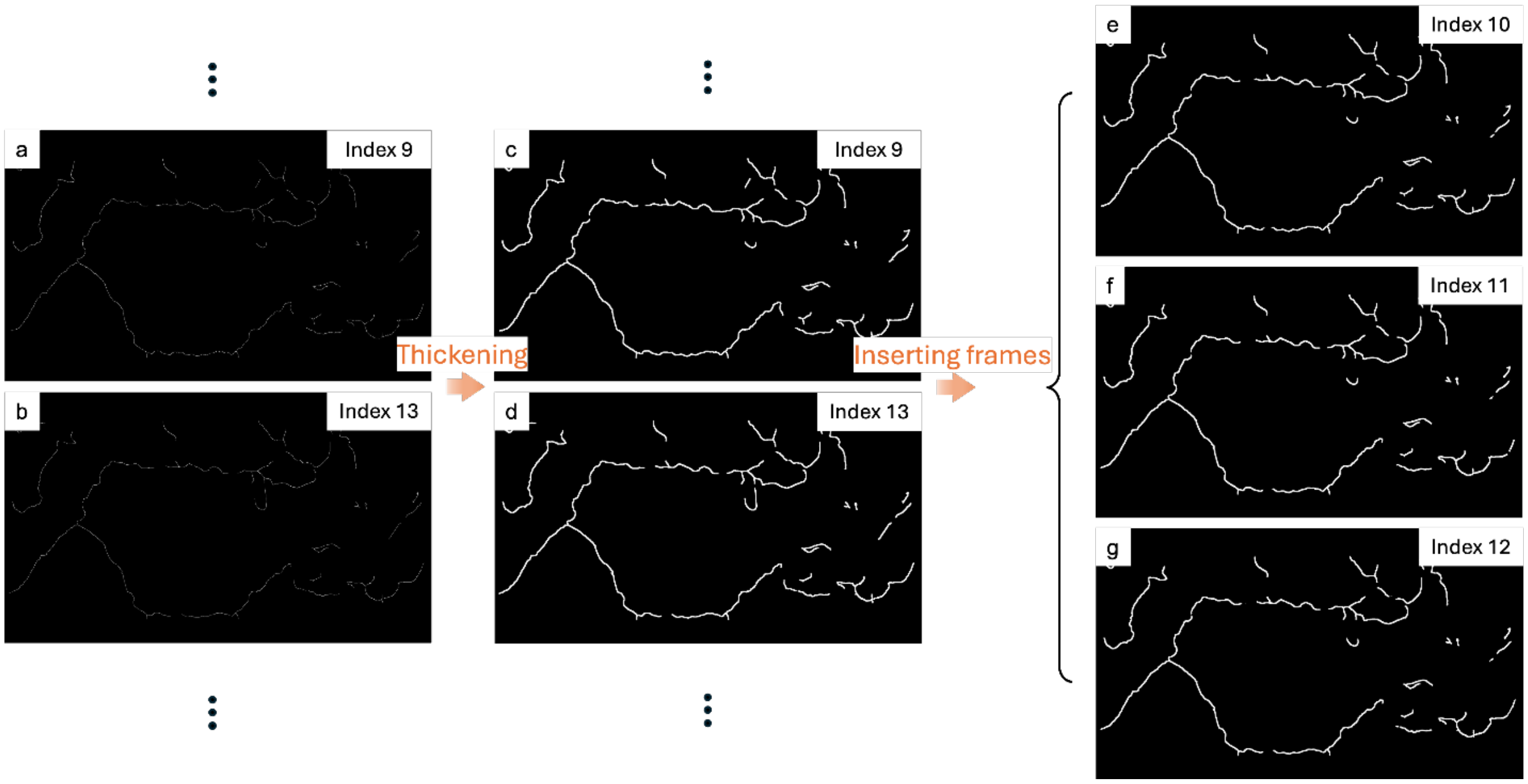

**Supplementary Fig. 15** | GB line thickening and insertion of intermediate frames. (a-b) Denoised GB line images of indices 9 and 13. (c-d) Thickened GB line images of indices 9 and 13 for improving visualization in 3D reconstruction. (e-g) Intermediate frames (indices 10, 11, 12, respectively) inserted between two GB line images shown in (c) and (d). Intermediate frames were generated using a signed distance transform–based interpolation of binary GB line images.

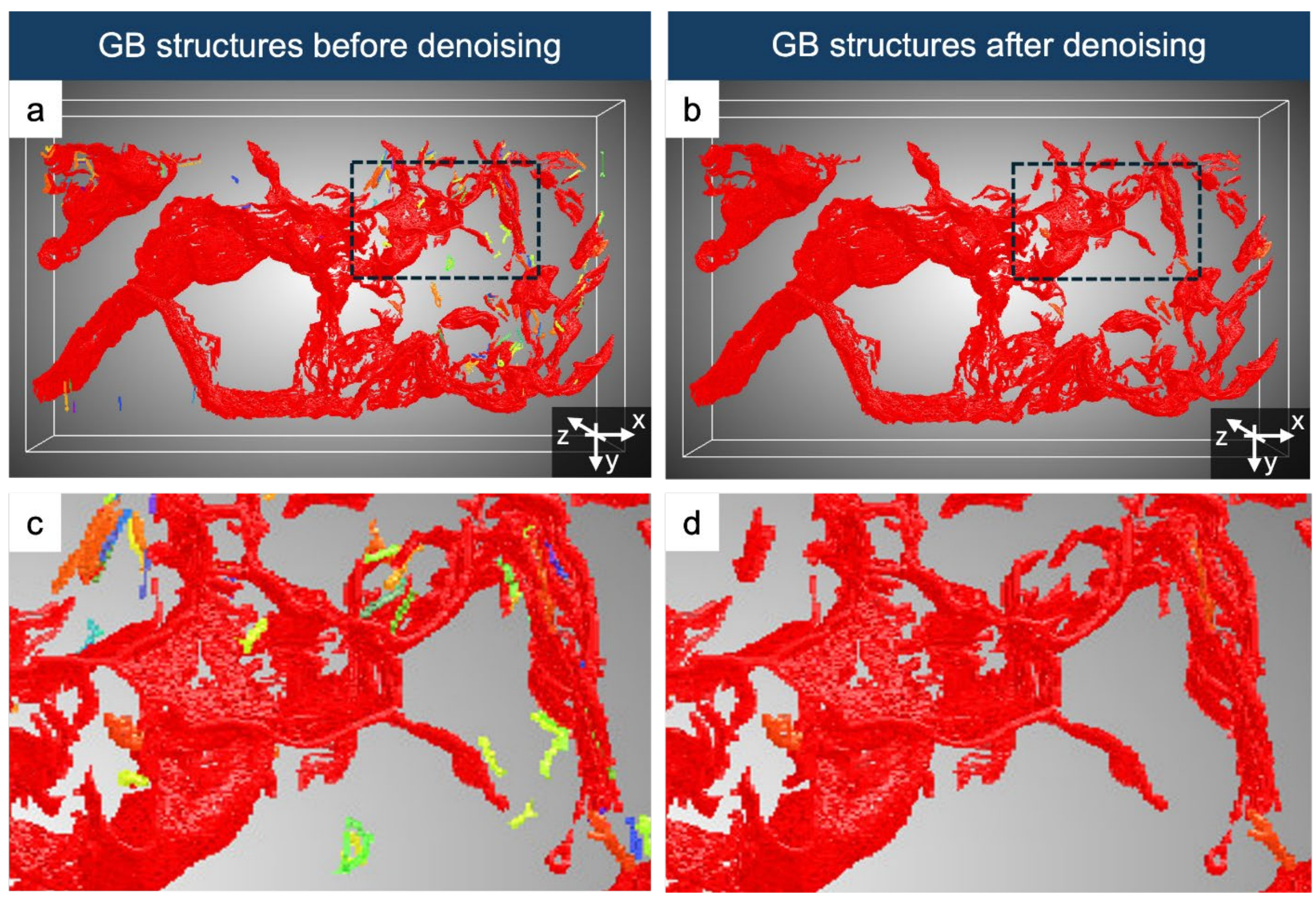

**Supplementary Fig. 16 |** Denoising of 3D structure of GBs. (a) Reconstructed 3D structure of GBs. (b) Denoised 3D structure of (a). Noise was removed using a connected-component–based volume filtering approach, in which small isolated volumes below a threshold were eliminated. (c–d) Local magnified views of the regions highlighted by the black boxes in (a) and (b), respectively.

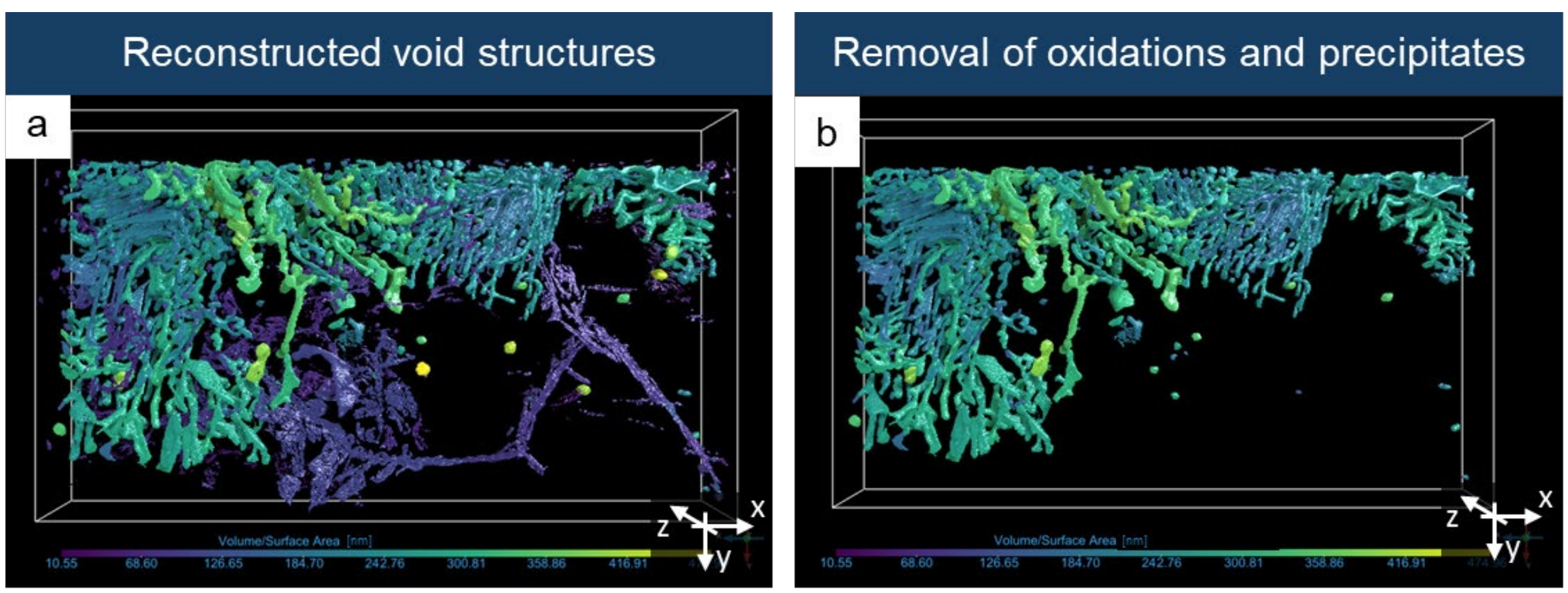

**Supplementary Fig. 17 |** Cleaning of oxidations and precipitates by filtering specific surface area of voids. (a) Reconstructed void structures. (b) Void structures after removal of oxidations and precipitates from (a) using specific surface area filtering.

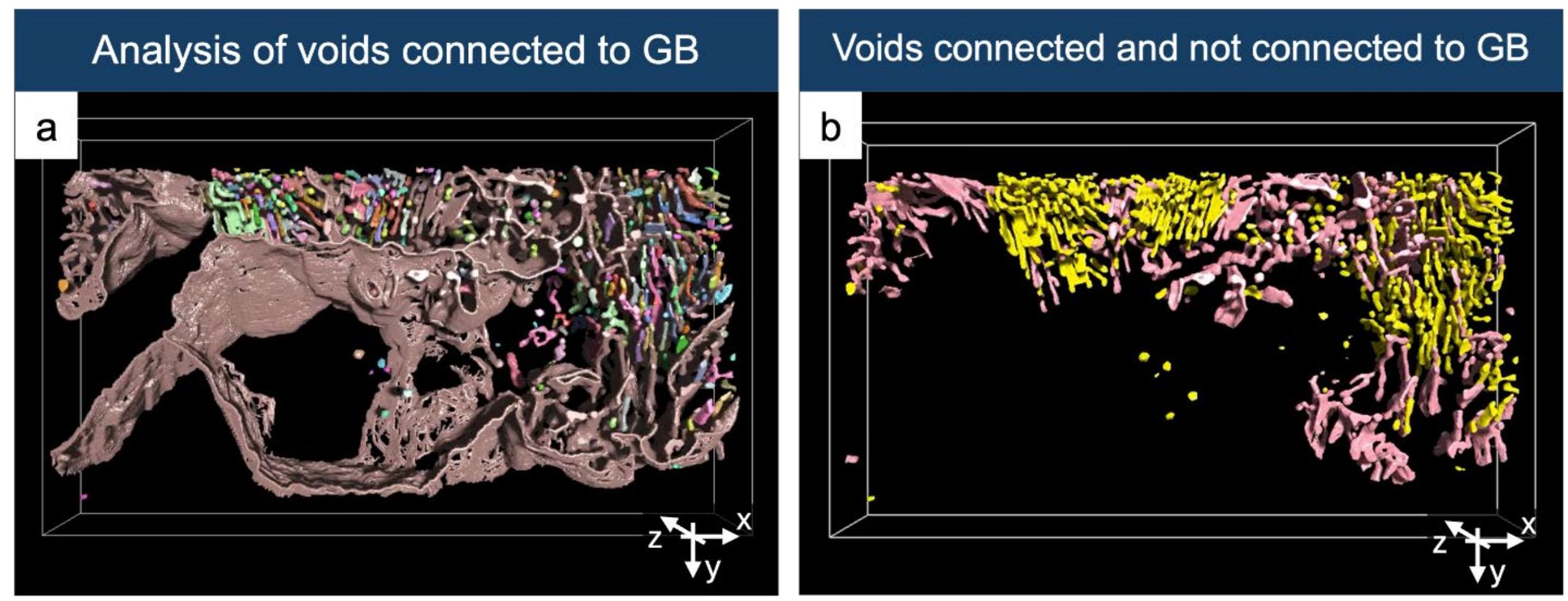

**Supplementary Fig. 18 |** Analysis of voids connected to GBs. (a) Connected component analysis of GBs and voids. Brown voids are connected to GBs based on the connected component analysis in Dragonfly. Voids of other colors are not connected to GBs. (b) Voids connected to GBs (red) and voids not connected to GBs (yellow).

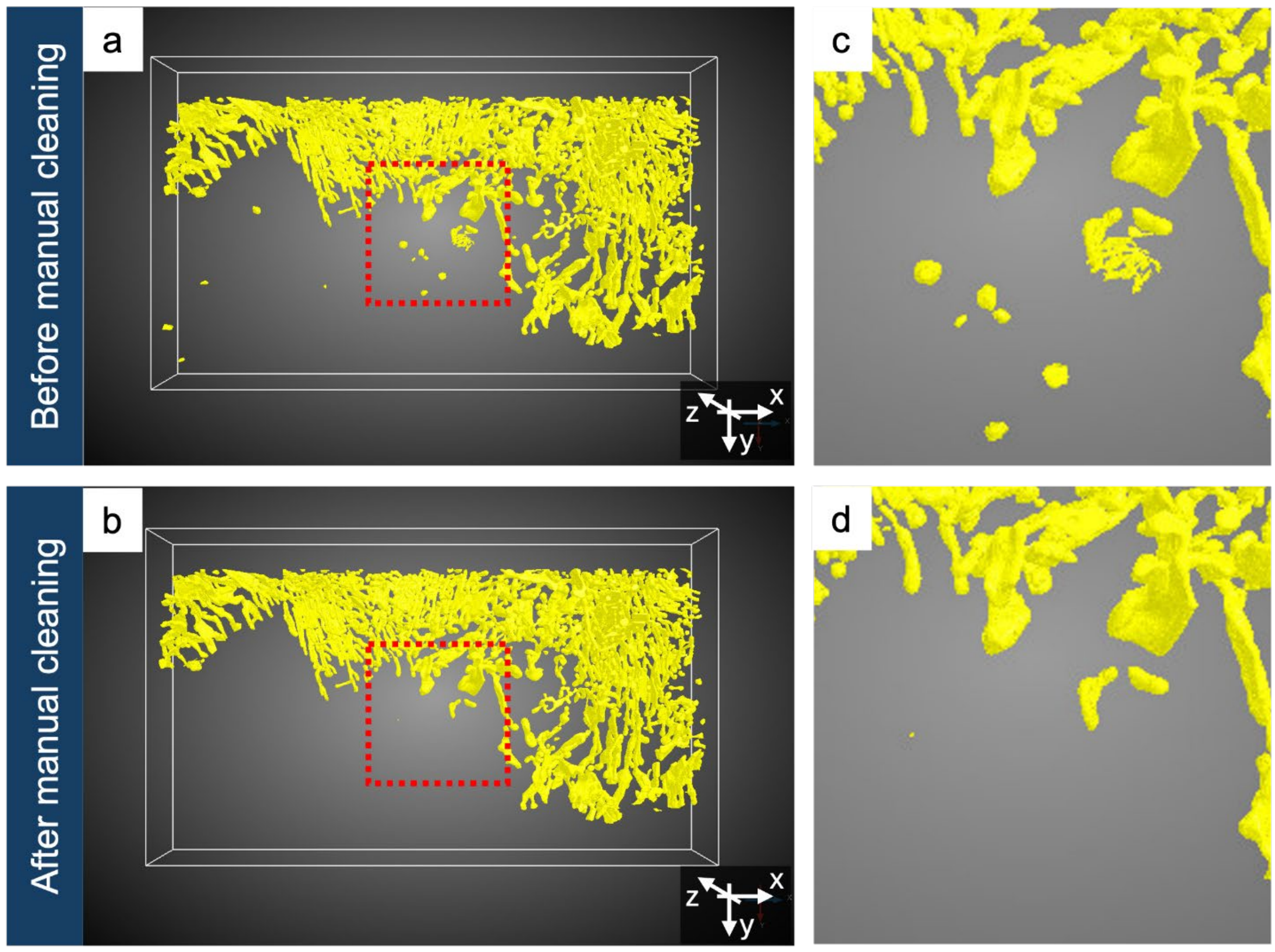

**Supplementary Fig. 19 |** Manual cleaning of oxides and precipitates (a) Void structures before manual cleaning. (b) Void structures after manual cleaning of oxides and precipitates from (a). (c–d) Local magnified views of the regions highlighted by the red boxes in (a) and (b), respectively.

## Captions for Supplementary Movies

**Supplementary Video 1:**

3D slice view of BSE images from 3D FIB-SEM tomography of a Ni-16.6 Cr sample after 8 h exposure to FLiNaK + 5 wt. % $EuF_3$ salt at 800 °C.

**Supplementary Video 2:**

3D slice view of IPFZ maps acquired from 3D FIB-SEM tomography of a Ni-16.6 Cr sample after 8 h exposure to FLiNaK + 5 wt. % $EuF_3$ salt at 800 °C.

**Supplementary Video 3:**

Reconstructed 3D void and GB structures. Voids connected to GBs are shown in red, and voids not connected to GBs are shown in yellow.

**Supplementary Video 4:**

Reconstructed 3D void structures. Voids connected to GBs are shown in red, and voids not connected to GBs are shown in yellow.

**Supplementary Video 5:**

A local region of the reconstructed 3D void and GB structures.

**Supplementary Video 6:**

Cross-sectional slice view of the reconstructed 3D void and GB structures.

**Supplementary Video 7:**

Single void 3D structure corresponding to the void shown in **Fig. 3b, c**.

**Supplementary Video 8:**

Single void 3D structure corresponding to the void shown in **Fig. 3e, f**.

**Supplementary Video 9:**

Single void 3D structure corresponding to the void shown in **Fig. 4b, c**.

**Supplementary Video 10:**

Single void 3D structure corresponding to the void shown in **Fig. 4e, f**.